\documentclass[preprint,pre,showpacs,floatfix,preprintnumbers,amsmath,amssymb]{revtex4}

\usepackage{graphicx}% Include figure files
\usepackage{dcolumn}% Align table columns on decimal point
\usepackage{bm}% bold math

\begin{document}

\title{Contact of Single Asperities with Varying Adhesion: Comparing
Continuum Mechanics to Atomistic Simulations}

\author{Binquan Luan and Mark~O. Robbins}

\affiliation{Department of Physics and Astronomy, The Johns Hopkins
 University, 3400 N.~Charles Street, Baltimore, Maryland 21218}

\date{March 22, 2006}

\begin{abstract}
Atomistic simulations are used to test the equations of continuum contact mechanics in nanometer scale contacts.
Nominally spherical tips, made by bending crystals or cutting crystalline or amorphous solids, are pressed into a flat, elastic substrate.
The normal displacement, contact radius, stress distribution, friction and lateral stiffness are examined as a function of load and adhesion.
The atomic scale roughness present on any tip made of discrete atoms is shown to have profound effects on the results.
Contact areas, local stresses, and the work of adhesion change by factors of two to four, and the friction and lateral stiffness vary by orders of magnitude.
The microscopic factors responsible for these changes are discussed.
The results are also used to test methods for analyzing experimental data with continuum theory to determine information, such as contact area, that can not
be measured directly in nanometer scale contacts.
Even when the data appear to be fit by continuum theory, extracted quantities can differ substantially from their true values.
\end{abstract} 


\maketitle

%\flushleft
\section{Introduction}

There has been rapidly growing interest in the behavior of materials at nanometer scales \cite{nanotechnology01}.
One motivation is to construct ever smaller machines \cite{bhushan04}, and a second is to improve material properties by controlling their structure at nanometer scales \cite{valiev02}.
For example, decreasing crystallite size may increase yield strength by suppressing dislocation plasticity, and material properties may be altered near free interfaces or grain boundaries.
To make progress, this research area requires experimental tools for characterizing nanoscale properties.
Theoretical models are also needed both to interpret experiments and to allow new ideas to be evaluated.

One common approach for measuring local properties is to press tips with characteristic radii of 10 to 1000 nm into surfaces using an atomic force microscope (AFM) or nanoindenter \cite{carpick97,carpick97b,carpick04,carpick04b,pietrement01,lantz97,schwarz97,schwarz97b,asif01,jarvis93,kiely98,wahl98}.
Mechanical properties are then extracted from the measured forces and displacements using classic results from continuum mechanics \cite{johnson85}.
A potential problem with this approach is that continuum theories make two key assumptions that must fail as the size of contacting regions approaches atomic dimensions.
One is to replace the atomic structure in the bulk of the solid bodies by a continuous medium with internal stresses determined by a continuously varying strain field.
The second is to model interfaces by continuous, differentiable surface heights with interactions depending only on the surface separation.
Most authors go further and approximate the contacting bodies by smooth spheres.

In a recent paper \cite{luan05}, we analyzed the limits of continuum mechanics in describing nanometer scale contacts between non-adhesive surfaces with curvature typical of experimental probes.
As in studies of other geometries \cite{miller96,landman96,vafek99}, we found that behavior in the bulk could be described by continuum mechanics down to lengths as small as two or three atomic diameters.
However, the atomic structure of surfaces had profound consequences for much larger contacts.
In particular, atomic-scale changes in the configuration of atoms on
nominally cylindrical or spherical surfaces produced factor of two changes in
the width of the contacting region and the stress needed to produce plastic yield, and order of magnitude changes in friction and stiffness.

In this paper we briefly revisit non-adhesive contacts with an emphasis on the role of surface roughness.
We then extend our atomistic studies to the more common case of adhesive interactions.
One important result is that the work of adhesion is very sensitive to small changes in the positions of surface atoms.
Changes in other quantities generally mirror those for non-adhesive tips, and small differences in the magnitude of these effects can be understood from geometrical considerations.
The results are used to test continuum-based methods of analyzing AFM measurements of friction and stiffness \cite{carpick97,carpick97b,carpick04,carpick04b,pietrement01,lantz97,schwarz97,schwarz97b}.
We show that the models may appear to provide a reasonable description when limited information about the true contact structure is available.
When the full range of information accessible to simulations is examined, one finds that the contact area and pressure distributions may be very different than inferred from the models.

Section \ref{sec:continuum} reviews continuum results for contact without and with adhesion, and briefly describes the effect of surface roughness.
The methods used in our atomistic simulations  and the geometries of the tips are described in Sec. \ref{sec:method}.
Section \ref{sec:nonadhere} presents results for purely repulsive interactions and Sec. \ref{sec:adhere} describes trends with the strength of adhesion.
A summary and conclusions are presented in Sec. \ref{sec:conclusions}.

\section{Continuum Contact Mechanics}
\label{sec:continuum}

As noted above, contact mechanics calculations assume that the contacting solids are described by continuum elasticity so that the discrete atomic structure can be ignored.
In most cases the two solids are also assumed to be isotropic with Young's moduli $E_1$ and $E_2$ and Poisson ratios $\nu_1$ and $\nu_2$.
Then the results depend only on an effective modulus $E^*$ satisfying:
\begin{equation}
1/E^*\equiv(1-\nu_1^2)/E_1 +(1-\nu_2^2)/E_2  .
\label{eq:effectmod}
\end{equation}
Three-dimensional crystalline solids are not isotropic, but the theories can still be applied with an effective $E^*$ that depends on orientation and is determined numerically \cite{johnson85}.

Continuum theories also neglect the atomic structure of the surface.
In most cases the surfaces are assumed to be spherical, with radii $R_1$ and $R_2$.
For elastic, frictionless solids the contact of two spheres is equivalent to contact between a sphere of radius $R = (R_1^{-1}+R_2^{-1})^{-1}$ and a flat solid \cite{johnson85}.
From Eq. (\ref{eq:effectmod}), one may then map contact between any two spherical surfaces onto contact between a rigid sphere of radius $R$ and a flat
elastic solid of modulus $E^*$.
This is the case considered in our simulations, and previous results indicate this mapping remains approximately correct at atomic scales \cite{luan05}.

Non-adhesive contact is described by Hertz theory \cite{johnson85}, which assumes solids interact with an infinitely sharp and purely repulsive ``hard-wall'' interaction.
The surfaces contact in a circular region of radius $a$ that increases with the normal force or load $N$ pushing the surfaces together as \cite{johnson85}:
\begin{equation}
\label{eq:hertza}
a = \left( \frac{3NR}{4E^*} \right) ^{1/3}\ \   .
\end{equation}
The normal pressure $p$ within the contact has a simple quadratic dependence on the radial distance from the center $r$:
\begin{equation}
p(r)=\frac{2aE^*}{\pi{R}}\sqrt{1-\frac{r^2}{a^2}}\ \   ,
\end{equation}
and the surfaces separate slowly outside the contact.
The normal displacement of the tip $\delta$ is related to $a$ by:
\begin{equation}
\label{eq:hertzd}
\delta_H = a^2/R = (\frac{3NR}{4E^*})^{2/3}  \ \ ,
\end{equation}
where the subscript $H$ indicates the Hertz prediction and
$\delta_{H}=0$ corresponds to the first contact between tip and substrate.

Adhesion can be treated most simply in the opposite limits of very short-range interactions considered by Johnson, Kendall and Roberts (JKR) \cite{johnson71} and of infinite range interactions considered by Derjaguin, Muller and Toporov (DMT) \cite{derjaguin75}.
The strength of adhesion is measured by the work of adhesion per unit area $w$.
In DMT theory the attractive forces just produce an extra contribution to the normal force,
so that $N$ is replaced by $N + 2\pi w R$ in Eqs. (\ref{eq:hertza}) and (\ref{eq:hertzd}).
JKR theory treats the edge of the contact as a crack tip and calculates the stress by adding the crack and Hertz solutions.
The normal force in Eq. (\ref{eq:hertza}) is then replaced by $N+ 3\pi w R + \left[6\pi w R N + (3\pi w R)^2\right]^{1/2}$ and the equation for $\delta$ is modified (Sec. \ref{sec:adhereload}).
The two approaches lead to very different functional relations between $a$ and $N$.  For example, the contact radius goes to zero at pulloff for DMT theory, but remains finite for JKR.
They also predict different values of the pulloff force, $N_c$, where the surfaces separate.
The normalized pulloff force, $N_c/\pi w R$, is -3/2 in JKR theory and -2 for DMT.
Finally, the surfaces separate outside the contact with infinite slope in
JKR theory, and gradually in DMT theory.

The Maugis-Dugdale (M-D) model \cite{maugis92} provides a simple interpolation between the JKR and DMT limits.
The surfaces are assumed to have a hard-wall contact interaction that prevents any interpenetration, plus a constant attractive force per unit area, $\sigma_0$, that extends over a finite distance $h_0$.
The work of adhesion is just the integral of the attractive force, implying $\sigma_0 h_0 = w$.
The M-D model produces coupled equations for the contact pressure that can be solved to yield a relation between the load, normal displacement, and area.
As discussed further in Section \ref{sec:adhere}, the edge of the contact is broadened by the finite interaction range, making it useful to define three characteristic radii that converge to the JKR value for $a$ in the limit of short-range interactions.

Maugis introduced a transition parameter \cite{maugis92}
\begin{equation}
\lambda \equiv \left( \frac{9Rw^2 }{2\pi {E^*}^2 h_0^3} \right)^{1/3},
\end{equation}
that measures the ratio of the normal displacement at pulloff from JKR theory to the interaction range $h_0$.
Tabor \cite{tabor76} had previously defined a similar parameter, $\mu$, that is about 16\% smaller than $\lambda$ for typical interaction potentials \cite{johnson97}.
Johnson and Greenwood \cite{johnson97} have provided an adhesion map characterizing the range of $\lambda$ over which different models are valid.
For $\lambda > 5$ the interaction range is short and JKR theory is accurate, while DMT is accurate for $\lambda < 0.1$.
For most materials, both $h_0$ and the ratio $w/E^*$ are of order 1 nm.
The JKR limit is only reached by increasing $R$ to macroscopic dimensions of micrometers or larger.
JKR theory has been tested in experiments with centimeter scale radii using the surface force apparatus (SFA) \cite{horn87} and hemispherical elastomers \cite{Shull02,newby95}.
Scanning probe microscope tips typically have $R$ between 10 and 100 nm, and the value of $\lambda \sim 0.1$ to 1 lies between JKR and DMT limits \cite{carpick97}.
The same is true in our simulations, where $\lambda$ for adhesive
tips varies between 0.1 and 0.75.
For this reason we will compare our results to M-D theory below.
We also found it useful to use a simple interpolation scheme suggested by
Schwarz \cite{schwarz03}.
Both he and Carpick et al. \cite{carpick99} have proposed
formulae for the contact radius that interpolate smoothly between DMT and JKR.
These approaches have been attractive in analyzing experimental data because of their simple analytic forms.

No direct measurement of contact area has been possible in nanometer scale single asperity contacts.
Instead, the contact area has been determined by measurements of contact stiffness \cite{carpick97,carpick97b,carpick04,carpick04b,pietrement01,lantz97,johnson97,jarvis93,kiely98,wahl98,asif01}, conductance \cite{lantz97}, or friction \cite{lantz97,carpick97,carpick97b,carpick04,carpick04b,schwarz97,schwarz97b}.
The validity of these approaches is not clear \cite{luan05}, and will be tested below.
The stiffness against normal displacements of the surfaces can be determined from the derivative of $N$ with respect to $\delta$ in M-D theory.
The tangential stiffness $k$ is normally calculated by assuming friction prevents sliding at the interface, even though all theories described above assume zero friction in calculating the contact area.
With this assumption $k=8G^* a$, where $G^* $ is the effective bulk shear modulus.
Relating the friction to contact area requires assumptions about the friction law.
Many authors have assumed that the friction is proportional to area \cite{carpick04,carpick04b,carpick97b,carpick97,pietrement01,lantz97,schwarz97,schwarz97b}, but simulations \cite{luan05,wenning01,muser01prl} and experiments in larger
contacts \cite{gee90,berman98} show that this need not be the case.

The effect of surface roughness on contact has been considered within the
continuum framework \cite{johnson85}.
In general, results must be obtained numerically.
One key parameter is the ratio of the root mean squared (rms) roughness of the surface, $\Delta$, to the normal displacement $\delta$.
When $\Delta/ \delta  < 0.05 $, results for nonadhesive contacts lie within
a few percent of Hertz theory \cite{johnson85}.
As $\Delta / \delta$ increases, the contact area broadens and the pressure in the central region decreases. 
Adhesion is more sensitive to roughness \cite{fuller75}.
The analysis is complicated by the fact that
$\Delta$ usually depends on the range of lengths over which it is measured.
The natural upper bound corresponds to the contact diameter and increases with
load, while the lower bound at atomic scales is unclear.
The role of roughness is discussed further in Secs. \ref{sec:nonadhere} and \ref{sec:adhere}.

\section{Simulation methods}
\label{sec:method}

We consider contact between a rigid spherical tip and an elastic substrate with effective modulus $E^*$.
As noted above, continuum theory predicts that this problem is equivalent to contact between two elastic bodies, and we found this equivalence was fairly accurate in previous studies of non-adhesive contact \cite{luan05}.
To ensure that any deviations from the continuum theories described above
are associated only with atomic structure, the substrate is perfectly elastic.
Continuum theories make no assumptions about the nature of the atomic
structure and interactions within the solids.
Thus any geometry and interaction potentials can be used to explore the
type of deviations from continuum theory that may be produced by atomic structure.
We use a flat crystalline substrate to minimize surface roughness,
and use tips with the minimum roughness consistent with atomic structure.
The interactions are simple pair potentials that are widely used in
studies that explore generic behavior \cite{allen87}.
They illustrate the type of deviations from continuum theory that
may be expected, but the magnitude of deviations for real materials
will depend on their exact geometry and interactions.

Atoms are placed on sites of a face-centered cubic (fcc) crystal with a (001) surface.
We define a characteristic length $\sigma$ so that the volume per atom is $\sigma^3$ and the nearest-neighbor spacing is $2^{1/6}\ \sigma$.
Nearest-neighbors are coupled by ideal Hookean springs with spring constant $\kappa$.
Periodic boundary conditions are applied along the surface of the substrate with period $L$ in each direction.
The substrate has a finite depth $D$ and the bottom is held fixed.
For the results shown below, $L=190.5\ \sigma$ and $D=189.3\ \sigma$.
The continuum theories assume a semi-infinite substrate, and we considered smaller and larger $L$ and $D$ to evaluate their effect on calculated quantities.
Finite-size corrections for the contact radius and lateral stiffness are negligible for $a/D < 0.1$ \cite{johnson85}, which covers the relevant range of $a/R < 0.2$.
Corrections to the normal displacement are large enough to affect plotted values.
We found that the leading analytic
corrections \cite{johnson01,sridhar04,adams06} were sufficient to
fit our results at large loads, as discussed in Sec. \ref{sec:adhereload}.
Note that previous simulations of AFM contact have used much shallower substrates ($D\sim 10\ \sigma$)
\cite{harrison99,landman90,sorensen96,nieminen92,raffi-tabar92,tomagnini93}.
This places them in a very different limit than continuum theories, although they provide interesting insight into local atomic rearrangements.

Three examples of atomic realizations of spherical tips are shown in Fig. \ref{fig:tips}.
All are identical from the continuum perspective, deviating from a perfect sphere by at most $\sigma$.
The smoothest one is a slab of f.c.c. crystal bent into a sphere.
The amorphous and stepped tips were obtained by cutting spheres from a bulk glass or crystal, and are probably more typical of AFM tips \cite{carpick97,foot1}.
Results for crystalline tips are very sensitive to the ratio $\eta$ between their nearest-neighbor spacing and that of the substrate, as well as their crystalline alignment \cite{muser01prl,wenning01}.
We will contrast results for an aligned commensurate tip with $\eta=1$ to those for an incommensurate tip where $\eta=0.94437$.
To mimic the perfectly smooth surfaces assumed in continuum theory, we also show results for a high density tip with $\eta=0.05$.
In all cases $R=100\ \sigma \sim 30$ nm, which is a typical value for AFM tips.
Results for larger radius show bigger absolute deviations from continuum predictions, but smaller fractional deviations \cite{luan05}. 

Atoms on the tip interact with the top layer of substrate atoms via a truncated Lennard-Jones (LJ) potential \cite{allen87}
\begin{equation}
V_{LJ}=-4\epsilon_i \left[ \left(\frac{\sigma}{r}\right)^6- \left(\frac{\sigma}{r}\right)^{12} \right] -V_{\rm cut},\qquad              r<r_{\rm cut}
\label{eq:lj}
\end{equation}
where $\epsilon_i$ characterizes the adhesive binding energy,
the potential vanishes for $r >r_{\rm cut}$, and the constant $V_{\rm cut}$ is subtracted so that
the potential is continuous at $r_{\rm cut}$.
Purely repulsive interactions are created by truncating the LJ
potential at its minimum $r_{\rm cut} = 2^{1/6}\ \sigma$.
Studies of adhesion use $r_{\rm cut}=1.5\ \sigma$ or $r_{\rm cut}=2.2\ \sigma$
to explore the effect of the range of the potential.

In order to compare the effective strength of adhesive interactions and the cohesive energy of the solid substrate, we introduce a unit of energy $\epsilon$ defined so that the spring constant between substrate atoms
$\kappa= 50\ \epsilon /\sigma^2$.
If the solid atoms interacted with a truncated LJ potential with $\epsilon$ and $r_{\rm cut}=1.5\ \sigma$, they would have the same equilibrium lattice constant
and nearly the same spring constant, $\kappa=57\ \epsilon/\sigma^2$, at low temperatures and small deformations.
Thus $\epsilon_i/\epsilon$ is approximately equal to the ratio of the interfacial binding energy to the cohesive binding energy in the substrate.

The elastic properties of the substrate are not isotropic.
We measure an effective modulus $E^* = 55.0\ \epsilon/\sigma^3$ for our
geometry using Hertzian contact of a high density tip.
This is between the values calculated from the Young's moduli in
different directions.
The sound velocity is also anisotropic.
We find longitudinal sound velocities of 8.5 and 9.5 $\sigma/t_{LJ}$
and shear velocities of 5.2 and 5.7 $\sigma/t_{LJ}$ along the (001) and
(111) directions, respectively.
Here $t_{LJ}$ is the natural characteristic time unit, $t_{LJ}=\sqrt{m\sigma^2/\epsilon}$, where $m$ is the mass of each substrate atom.
The effective shear modulus for lateral tip displacements is $G^* = 18.3\ \epsilon/\sigma^3$.

The simulations were run with the Large-scale Atomic/Molecular Massively Parallel Simulator (LAMMPS) code \cite{plimpton95,lammps}.
The equations of motion were integrated using the velocity-Verlet algorithm
with time step 0.005$\ t_{LJ}$ \cite{allen87}.
Temperature, $T$, only enters continuum theory through its effect on constitutive properties, and it is convenient to run simulations at low temperatures to minimize fluctuations.
A Langevin thermostat was applied to solid atoms to maintain $T$=0.0001$\ \epsilon/k_B$, where $k_B$ is Boltzmann's constant.
This is about four orders of magnitude below the melting temperature of
a Lennard-Jones solid.
The damping rate was $0.1\ t_{LJ}^{-1}$, and damping was only applied perpendicular to the sliding direction in friction measurements.

In simulations, a tip was initially placed about $r_{\rm cut}$ above the substrate.
It was then brought into contact by applying a very small load, with the lateral position kept fixed.
The load was changed in discrete steps and the system was allowed to equilibrate for 350$\ t_{LJ}$ at each load before making measurements.
This interval is about 20 times longer than the time for sound to propagate across the substrate, and allowed full stress equilibration.
Results from loading and unloading cycles showed no noticeable hysteresis.
To obtain results near the pulloff force, we moved the tip away from the substrate at a very slow velocity $v=0.0003\ \sigma/t_{LJ}$ and averaged results over small ranges of displacement.
This approach was consistent with constant load measurements and allowed us to reach the region that is unstable at constant load.

To compare to continuum predictions we calculated the stresses
exerted on the substrate by the tip.
The force from the tip on each substrate atom was divided by the
area per atom to get the local stresses.
These were then decomposed into a normal stress or pressure $p$
on the substrate,
and a tangential or shear stress ${\bf \tau}_{\rm sur}$.
The continuum theories described in Sec. \ref{sec:continuum} assume that
the projection of the force normal to the undeformed substrate
equals the projection normal to the locally deformed substrate.
This is valid
in the assumed limits of $a/R << 1$ and ${\bf \tau}_{\rm sur}=0$.
It is also valid for most of our simulations (within $<2$\%),
but not for the case of bent commensurate tips where ${\bf \tau}_{\rm sur}$
becomes significant.
Normal and tangential stresses for
bent commensurate tips were obtained using the local surface orientation
of the nominally spherical tip.
Correcting for the orientation changed the normal stress by less
than 5\% of the peak value, and the shear stress
by less than 20\%.

Friction forces are known to vary with many parameters \cite{muser04}.
Of particular concern is the dependence on extrinsic quantities such as the stiffness of the system that imposes lateral motion.
Results at constant normal load are often very different than those at fixed height, motion perpendicular to the nominal sliding direction can raise or lower friction, and the kinetic friction can be almost completely eliminated in very stiff systems \cite{muser03acp,socoliuc04}.
A full re-examination of these effects is beyond the scope of this paper.
Our sole goal is to measure the friction in a consistent manner that allows us to contrast the load dependent friction for different tip geometries and minimizes artifacts from system compliance.

In friction simulations, the tip is sheared at a constant low velocity $v'=0.01\ \sigma/t_{LJ}$ along the (100) direction with a constant normal load.
This is typical of AFM experiments where the low normal stiffness of the cantilever leads to a nearly constant normal load, and the high lateral stiffness limits lateral motion in the direction perpendicular to the sliding direction.
The measured friction force varies periodically with time as the tip moves by a lattice constant of the substrate.
The time-averaged or kinetic friction during sliding is very sensitive to both lateral stiffness and load \cite{socoliuc04}.
We focus instead on the peak force, which is less sensitive.
In the limit of low sliding velocities this would correspond to the static friction.
For bent and stepped commensurate tips there is a single strong friction peak.
For incommensurate and amorphous tips, there may be multiple peaks of different size corresponding to escape from different metastable energy minima \cite{muser01prl,muser03acp}. 
The static friction was determined from the highest of these friction peaks,
since lateral motion would stop at any lower force.
With a single peak per period, the time between peaks is
$\sim \sigma/v'=100\ t_{LJ}$.
This is several times the sound propagation time, and the measured force should be close to the static friction.
For incommensurate tips the time between peaks was an order of magnitude smaller and dynamic effects may be more significant.
However, they are not expected to affect the load dependence significantly, and are much too small to affect the dramatic difference between incommensurate and other tips.

The total lateral stiffness of the system, $k$, corresponds to the derivative of $F$ with lateral tip displacement evaluated at a potential energy minimum.
Since the tip is rigid, $k$ is determined by displacements in the substrate
and at the interface.
The interfacial stiffness $k_{\rm i}$ and substrate stiffness $k_{\rm sub}$
add in series because stress is transmitted through interfacial interactions
to the substrate.
Thus the total stiffness is \cite{socoliuc04,luan05}:
\begin{equation}
k^{-1}=k_{\rm sub}^{-1}+k_{\rm i}^{-1} \ \ .
\label{eq:stiff}
\end{equation}
If the tip were not rigid, it would also contribute a term to the
right-hand-side of Eq. (\ref{eq:stiff}).

We evaluate $k$ from the derivative of $F$ during sliding, making the assumption that the results are in the quasisatic limit.
For bent and stepped commensurate tips there is a single potential energy minima, and for amorphous tips one minimum dominated the periodic force.
For incommensurate tips, there are many closely spaced minima and we evaluate $k$ from the derivative in the minimum preceding the largest friction peak.
Due to the small magnitude of forces and short time intervals,
the relative errors in these values are as big as 50\%.
To estimate the lateral stiffness in the substrate, $k_{\rm sub}$, we fix the relative positions of those substrate atoms that lie inside the contact, and move them laterally at a slow velocity.
The total force between these atoms and the rest of the substrate is measured and its derivative with respect to distance gives the lateral stiffness in the substrate.  
In principal, there might also be some relative displacement between
atoms in the contact that is not captured by this approach, but
the results for the substrate stiffness are consistent with continuum
predictions.

Values of the adhesion energy per unit area $w$ were obtained for
flat, rigid surfaces of the same nominal geometry as the tip.
For bent crystal tips (Fig. \ref{fig:tips}),
the tip was just flattened back into a crystal.
For stepped tips, we used an unstepped crystal with the same spacing
and interactions.
For amorphous tips, an amorphous solid was cleaved with a flat
surface rather than a sphere.
The resulting surfaces were then brought into contact with
the substrate and allowed to equilibrate at zero load.
At the low temperatures used here, the adhesion energy
is just the potential energy difference between
contacting and separated configurations.

\section{Nonadhesive contacts}
\label{sec:nonadhere}
\subsection{Pressure distribution}

Figure \ref{fig:hertzpofr} contrasts the distribution of normal
pressure $p$
under five tips: (a) dense, (b) bent commensurate, (c) bent incommensurate, (d) amorphous and (e) stepped.
In each case, $R=100\ \sigma$ and the dimensionless load is $N/(R^2 E^*) =0.0018$.
Hertz theory predicts the same pressure distribution (solid lines) for all tips.
Points show the actual local pressure on each substrate atom as a function of radial distance $r$ from the center of the spherical tip, and circles in (c) and (d) show the average over bins of width $\sigma$.
Clearly, small deviations in atomic structure lead to large changes in the mean pressure and the magnitude of pressure fluctuations.
We find that these deviations become larger as $N$ is decreased, and the contact radius drops closer to the atomic size.

One possible source of deviations from Hertz theory is friction,
but we find the mean tangential forces are small in most cases.
The exception is the bent commensurate tip
(Fig. \ref{fig:hertzpofr}(b)), where the tangential stress rises
with $r$ and is comparable to the normal stress near the edge of the contact.
This result is not surprising given the high friction measured for commensurate
tips below, and reflects the strong tendency for atoms in the substrate to
remain locked in epitaxial registry with atoms in the tip.
However, the deviation from Hertz theory is in the opposite direction
from that expected from friction.
Since this contact was made by gradually increasing the load, friction
should decrease the contact size rather than broadening it.

Another possible origin of the deviations from Herts theory is surface
roughness.
From continuum theory (Sec. \ref{sec:continuum}), this is characterized by the ratio of rms surface roughness $\Delta$ to normal displacement $\delta$.
The normal displacement for all tips is about the same, $\delta \approx 1.5\ \sigma$, but $\Delta$ is difficult to define. 
The reason is that there is no unique definition of the surface height for a given set of atomic positions.
For example, one might conclude that $\Delta=0$ for the substrate, since all atoms lie on the same plane.
However, if a tip atom were moved over the surface with a small load, its height would increase as it moved over substrate atoms and be lowest at sites centered between four substrate atoms \cite{muser03acp}.
For the parameters used here, the total height change is about 0.33$\ \sigma$.
Similar deviations from a sphere are obtained for the bent commensurate and incommensurate tips.
The height change decreases as the ratio of the nearest-neighbor spacing to the Lennard-Jones diameter for interfacial interactions decreases,
and is only 0.0007$\ \sigma$ for the dense tip.
Amorphous and stepped tips have additional roughness associated with variations in the position of atomic centers relative to a sphere.
The total variation is about $\sigma$, or about three times the height change as an atom moves over the substrate.
A reasonable estimate is that $\Delta/\delta < 0.1$ for the bent commensurate and incommensurate tips, $\Delta/\delta < 10^{-3}$ for the dense tip, and $\Delta/\delta \sim 0.3$ for the amorphous and stepped tips.
However, the ambiguity in $\Delta$ is one of the difficulties in applying continuum theory in nanoscale contacts.

The closely spaced atoms on the dense tip approximate a continuous sphere, and the resulting pressure distribution is very close to Hertz theory (Fig. \ref{fig:hertzpofr}(a)).
Results for the bent commensurate tip are slightly farther from Hertz theory.
The deviations can not be attributed to roughness, because fluctuations at a given $r$ are small, and the pressure in the central region is not decreased.
The main change is to smear the predicted sharp pressure drop at the edge of the contact.
This can be attributed to the finite range of the repulsive potential between surfaces.

We can estimate the effective interaction range by the change in height of an atom, $dh = 0.04\ \sigma$, as $p/E^*$ decreases from 0.1 to 0.
The effective range is much smaller for the dense tip because $\sim 400$ times as many atoms contribute to the repulsive potential.
In Hertz theory \cite{johnson85}, the separation between surfaces only increases with distance $(r-a)$ from the edge of the contact as $(8/3\pi)(r-a)^{3/2} (2a)^{1/2}/R$.
Equating this to $dh$ gives $r-a \approx 1\ \sigma$ for the bent commensurate tip, which is very close to the range over which the edge of the contact is shifted from the Hertz prediction.
Note that this analysis predicts that the shift in the edge of the contact will grow as $\sqrt{R}$, and simulations with larger $R$ confirm this.
However, the fractional change in $a$ decreases as $1/\sqrt{R}$.
The larger values of pressure at low $r$ result from the greater stiffness of the repulsive potential as $p$ increases.
All of the above effects could be included in continuum theory by changing the form of the repulsive potential \cite{greenwood97}.

For bent incommensurate and amorphous tips (Fig. \ref{fig:hertzpofr} (c) and (d)), the variations in pressure at a given $r$ are as large as the mean (circles)
\cite{footn}.
While all atoms on the commensurate tip can simultaneously fit between substrate atoms, minimizing fluctuations in force, atoms on the incommensurate tip sample all lateral positions and experience a range of forces at a given height.
The mean pressure for the incommensurate tip remains fairly close to the commensurate results, but there is slightly more smearing at large $r$ due to the variations in relative height of substrate and tip atoms.
The mean pressure on the amorphous tip shows the depression at small $r$ and increase at large $r$ that are expected for rough tips in continuum theory \cite{johnson85}.
The magnitude of the central drop is about 18\%, which is consistent with $\Delta/\delta \sim 0.2$ in continuum theory (Fig. 13.12 of Ref. \cite{johnson85}).
The lack of a noticeable drop for incommensurate tips implies that the effective
$\Delta/\delta < .03$.
The implication is that the incoherent height changes on amorphous tips contribute to the effective roughness in continuum theory, while the atomic corrugation on bent tips does not.
The effective roughness in both cases is about 0.1$\ \sigma$ smaller than our estimates for $\Delta$ above.

Results for stepped tips show the largest deviations from Hertz theory,
and they are qualitatively different than those produced by random roughness.
The terraced geometry of this tip (Fig. \ref{fig:tips}) is closest to that of a flat punch.
In continuum theory, the pressure on a flat punch is smallest in the center, and diverges as the inverse square root of the distance from the edge.
The simulation results show qualitatively similar behavior.
The main effect of atomic structure is to cut off the singularity at a distance corresponding to an atomic separation.
Similar effects are observed in simulations of other geometries \cite{vafek99,miller96,landman96}.
Note that the terraces are only flat because the sphere was cut from a crystal that was aligned with the substrate.
We also examined tips cut from a crystal that was slightly tilted away from the (001) direction \cite{foot1}.
This produces inclined terraces that contact first along one edge.
The resulting pressure distribution is very different, and closest to the continuum solution for contact by an asymmetric wedge.

Figure \ref{fig:hertzpofr} has an important general implication about the probability $P(p)$ of finding a local pressure $p$ at a point in the contact.
For smoothly curved surfaces, continuum theory predicts that the derivative of the pressure diverges at the edge of the contact \cite{johnson85}.
Thus $P(p) \rightarrow 0$ as $p\rightarrow 0$ \cite{persson01}.
The finite resolution at atomic scales always smears out the change in
$p$, leading to a non-zero value of $P(0)$.
Indeed, the approximately constant value of $dp/dr$ near the contact edge in Fig. \ref{fig:hertzpofr} leads to a local maximum in $P$ at $p=0$.
Similar behavior is observed for randomly rough atomic contacts \cite{luan05mrs} and in continuum calculations for piecewise planar surfaces \cite{hyun04,pei05}.

Plastic deformation is usually assumed to occur when the deviatoric
shear stress $\tau_s$ exceeds the yield stress of the material.
In Hertz theory, $\tau_s$ reaches a maximum value at a depth of about 0.5$a$.
The pressure variations at the surface shown in Fig. \ref{fig:hertzpofr} lead to changes in both the magnitude and position of the peak shear stress \cite{luan05}.
Factors of two or more are typical for amorphous and stepped tips.
Thus tip geometry may have a significant impact on the onset of yield.
Of course atomistic effects also influence the yield stress at nanometer scales, and a full evaluation of this effect is left to future work.
Saint-Venant's principal implies that the pressure distribution should become independent of tip geometry at depths greater than about $3a$, but the shear stress at these depths is substantially smaller than peak values and yield is unlikely to occur there.

\subsection{Variations with load}

Figure \ref{fig:hertz} shows the load dependence of (a) normal displacement, (b) radius, (c) friction and (d) lateral stiffness for the same tips as Fig. \ref{fig:hertzpofr}.
Each quantity is raised to a power that is chosen so that Hertz theory predicts the indicated straight line.
A small finite-depth correction ($\sim 2$\%) is applied to the Hertz prediction for $\delta$ (Eq. (\ref{eq:depth})).

As also found for cylindrical contacts \cite{luan05}, the normal displacement shows the smallest deviation from Hertz theory because it represents a mean response of many substrate atoms.
Results for all bent crystals are nearly indistinguishable from the straight line.
Results for the stepped surface are lower at small loads.
Since the entire tip bottom contacts simultaneously, it takes a larger load to push the tip into the substrate.
The amorphous results are shifted upwards by a fairly constant distance of about 0.2 $\sigma$.
We define the
origin of $\delta$ as the tip height where the first substrate atom exerts a repulsive force on the tip.
This is strongly dependent on the height of the lowest tip atom, while subsequent deformation is controlled by the mean tip surface.
Agreement with Hertz is greatly improved by shifting the amorphous curve by this small height difference.
Note that the zero of $\delta$ is difficult to determine experimentally and is usually taken as a fit parameter.
If this is done, even results for the amorphous system yield values of $R$ and $E^*$ that are within 10\% of their true values.
Thus measurements of $\delta$, interpreted with continuum theory for spheres, can provide useful values of elastic properties at nanometer scales.

As expected from the observed pressure distributions (Fig. \ref{fig:hertzpofr}), the contact radius is generally larger than the Hertz prediction.
The shift is smallest for the dense tip because it approximates a continuous surface and the high density leads to a repulsive potential that rises more than a hundred times more rapidly than for other tips.
Results for bent crystal and amorphous tips are shifted upwards by fairly load-independent offsets of $\sim 1 - 3\  \sigma$, leading to large fractional errors at low loads (up to 100\%).
The stepped crystal shows qualitatively different behavior, with $a$ rising in discrete steps as sequential terraces come into contact.
Note that the size of the first terrace is not unique, but depends on the registry between the bounding sphere and crystalline lattice \cite{foot1}.
Larger deviations may be observed when the first step has very few atoms.
Such tips may be more likely to be chosen for AFM studies because they tend to give sharper images.

In order to predict the friction between surfaces, one must make assumptions about how $F$ depends on area and load.
The straight line in Fig. \ref{fig:hertz}(c) corresponds to a friction force that is proportional to load.
Static friction values for bent and stepped commensurate surfaces are consistent with this line and a coefficient of friction $\mu \equiv F/N = 0.63$.
Analytic \cite{muser01prl} and simple numerical \cite{ringlein04} models show that this is a general feature of mated surfaces where each tip atom has the same lateral position relative to substrate atoms.
The friction on amorphous and incommensurate surfaces is always lower and scales more closely with the contact area, as indicated by broken line fits to
$F \propto N^{2/3}$ and discussed
further in Sec. \ref{sec:adhereload} \cite{foot2}.
Many authors have made this assumption in fitting experimental data, but it is not obvious why it should hold.
The friction per unit area between flat amorphous surfaces decreases as the square root of the area, but rises linearly with the normal pressure \cite{muser01prl}.
Wenning and M\"user have noted that these two factors may combine for spherical contacts to give a net friction that rises linearly with area \cite{wenning01}.
However, their argument would predict that the frictional force in a
cylinder-on-flat geometry would not scale with area, and our previous simulations found that it did \cite{luan05}.  
Continuum theory predicts that the lateral stiffness $k=8G^* a$, and should follow the straight line in Fig. \ref{fig:hertz}(d).
Measured values of the total stiffness (open symbols) are always substantially lower.
This is because continuum theory assumes that there is no displacement at the interface, only within the substrate.
In reality, the frictional force is always associated with atomic scale displacements of interfacial atoms relative to a local energy minimum \cite{ringlein04,muser03acp}.
The derivative of force with displacement corresponds to an interfacial stiffness $k_{\rm i}$ that adds in series with the substrate contribution (Eq. (\ref{eq:stiff})) \cite{socoliuc04,luan05}.
Our numerical results show that $k_{\rm i}$ can reduce $k$ by more than an order of magnitude, particularly for tips where $F$ is small.
We also evaluated the substrate stiffness $k_{\rm sub}$ by forcing all contacting substrate atoms to move together.
These results (filled symbols) lie much closer to the Hertz prediction.
Only the stepped tip shows large discrepancies, and these are correlated with the large deviation between the measured and predicted contact radii.

\section{Adhesive contacts}
\label{sec:adhere}

\subsection{Pressure distribution}

Figure \ref{fig:cutoff} compares the calculated pressure distribution in adhesive contacts with the Maugis-Dugdale prediction (lines).
A bent commensurate tip was used to minimize deviations from continuum predictions for a sphere.
Results for two different $r_{\rm cut}$ are presented to indicate difficulties
in fitting longer-range interactions to M-D theory.
The work of adhesion was calculated for unbent surfaces
(Sec. \ref{sec:method})
with $\epsilon_i/\epsilon = 0.5$, yielding $w=1.05\ \epsilon/\sigma^2$ and
$1.65\ \epsilon/\sigma^2$ for $r_{\rm cut}=1.5\ \sigma$ and $2.2\ \sigma$, respectively.
This leaves only one fitting parameter in M-D theory.
For the dashed lines,
the width of the attractive interaction $h_0 = w/\sigma_0$ was chosen to coincide with the range of the atomic potential.
The dotted line shows a fit with $h_0=0.8\ \sigma$ for $r_{\rm cut}=2.2\ \sigma$,
which gives better values for the pulloff force, but poorer radii
(Sec. \ref{sec:adhereload}).

In M-D theory, it is common to identify two radii, $a$ and $c$, with the inner and outer edges of the plateau in the pressure, respectively \cite{maugis92,johnson97}.
For $r< a$ the surfaces are in hard-sphere contact, and for $a<r<c$ they are separated and feel the constant attraction.
The continuously varying interactions between atoms in our simulations lead to qualitatively different behavior.
There is no sharp transition where the surfaces begin to separate, and the attraction shows a smooth rise to a peak, followed by a decay.
To facilitate comparison to continuum theories, we introduce the three characteristic radii indicated by arrows for each $r_{\rm cut}$.
The innermost, $r_a$, corresponds to the point where the interaction changes from repulsive to attractive and can be calculated in M-D theory.
The outermost, $r_c$, corresponds to $c$ in M-D theory -- the point where interactions end.
The middle, $r_b$, corresponds to the peak attraction.
It was also studied by Greenwood \cite{greenwood97} for contact of a featureless sphere that interacted with the flat substrate through Lennard-Jones interactions.
He found $r_b$ lay close to the JKR prediction for contact radius at large loads.
All three radii converge in the limit of repulsive interactions or the JKR limit of an infinitely narrow interaction range.

At small radii the atomistic results for $p$ lie above M-D theory, and they drop below near $r/\sigma=10$.
These deviations can be attributed to the increasing stiffness of the repulsion between tip and substrate atoms with increasing pressure.
Just as in the non-adhesive case, the stiffer interactions in the center of the tip lead to bigger pressures for a given interpenetration.
The change in pressure with separation produces less smearing at the edge of the repulsive region of the contact than for the non-adhesive case (Fig. \ref{fig:hertzpofr}(b)), and the values of $r_a$ are very close to M-D theory.
This can be understood from the fact that surfaces separate much more rapidly in JKR theory ($\propto (r-a)^{1/2}$) than in Hertz theory ($\propto (r-a)^{3/2}$) (Sec. \ref{sec:nonadhere}).
Thus the same height change in the repulsive region corresponds to a much smaller change in radius.
Of course the finite range of the attractive tail of the potential leads to a broad region of adhesive forces out to $r_c$. 
The continuous variation of $p$ in the attractive tail is only crudely approximated by M-D theory.
The difficulty in determining the optimum choice for $h_0$ increases with the range of interactions, as discussed further below.

Figure \ref{fig:r05pofr} shows the effect of tip geometry on pressure distribution.
We found that the work of adhesion was very sensitive to tip geometry.
To compensate for this effect, we varied $\epsilon_i$ (Table I) to yield the same $w$ for the tips in Fig. \ref{fig:r05pofr}.
Then all tips should have the same pressure distribution in continuum theory.
The M-D predictions for $p$ are not shown because even the bent commensurate tips produce significantly different results (Fig. \ref{fig:cutoff}).
Instead, we compare other tips to the bent commensurate tip.

As for non-adhesive tips, local fluctuations in pressure are small for commensurate tips (Fig. \ref{fig:r05pofr}(a) and (d)) and comparable to the mean pressure for incommensurate or amorphous tips (Fig. \ref{fig:r05pofr}(b) and (c)).
Note however, that the fluctuations become smaller in the adhesive regime (large $r$).
This is because the potential varies more slowly with height, so fluctuations in the separation between atoms have less effect on $p$.
One consequence is that the outer edge of the contact is nearly the same for commensurate and incommensurate tips.
The radii for the amorphous tip are significantly larger than bent tips, presumably because of a much greater effective roughness.
As for nonadhesive tips, the mean pressure on incommensurate tips is close to the commensurate results.
Adhesion reduces the roughness-induced drop in pressure in the central region of the amorphous tip.
For the stepped tip, the contact radius is dominated by the size of the terraces, but adhesive tails are visible outside the edge of each terrace.
Only $r_c$ is easily defined for the stepped tips.
This increases in a nearly stepwise manner, and its load dependence is not shown below.
For the amorphous and incommensurate tips, radii are determined from the mean pressure at a given radius (open circles).
Errorbars are less than 0.5 $\sigma$.

\subsection{Variations of radius and displacement with load}
\label{sec:adhereload}

Figure \ref{fig:r05rofN}(b) compares the measured contact radii for the tips of Fig. \ref{fig:r05pofr} to M-D theory as load is varied.
The simulation results for $r_c$ (open symbols) decrease with decreasing load as $r_a$ (closed symbols) decreases to zero.
All interactions in the contact are then adhesive.
As $r_c$ continues to drop, the area contributing to adhesion drops, and the load rises back toward zero.
This regime is not considered in the original M-D theory.
The extension to $r_a=0$ by Johnson and Greenwood (JG) \cite{johnson97} is shown by a dashed line in the figure.
Along this line the stress in the contact is constant,
giving $N=-\sigma_0 \, \pi  r_c^2$.

As for non-adhesive tips, the contacts tend to be larger than continuum theory predicts.
However, the shift for bent tips is smaller than in the non-adhesive case, and the commensurate and incommensurate results are closer, as expected from Fig. \ref{fig:r05pofr}.
Larger deviations are observed for the amorphous tip, with radii typically 2 or 3 $\sigma$ larger than predicted.
The deviation becomes most pronounced at negative loads, where the amorphous tip remains in contact well below the predicted pulloff force.

Figure \ref{fig:r05rofN}(a) compares the value of $r_b$ to the JKR prediction for contact radius.
As found by Greenwood \cite{greenwood97}, the numerical results are close to JKR at large loads, but deviate at negative loads because M-D and JKR predict different pulloff forces.
Since JKR assumes a singular adhesive stress at the radius of the contact, it seems natural that its predictions lie closest to the position of the peak tensile stress.

Figure \ref{fig:e05rofN} shows the calculated radii for bent and amorphous tips with the same interaction energy $\epsilon_i/\epsilon = 0.5$.
The small changes in tip geometry lead to a roughly four fold variation in both $w$ and $N_c$ (Table I).
The largest $w$ is obtained for commensurate tips because all atoms can optimize their binding coherently.
Atoms on incommensurate tips sample all positions relative to the substrate, and can not simultaneously optimize the binding energy.
The larger height fluctuations on amorphous tips lead to even greater reductions in $w$.

In the simulations, the pulloff force, $N_c$, corresponds to the most negative load where the surfaces remain in contact and the $r_i$ can be defined.
Its normalized value, $N_c/\pi w R$, is equal to -3/2 in JKR theory,  -2 in DMT theory, and lies between these values for M-D theory.
Table I shows normalized results for various tips.
As expected from the good agreement in Figs. \ref{fig:r05rofN} and \ref{fig:e05rofN}, results for bent tips lie between JKR and DMT values and can be fit with M-D theory.
The values for stepped and amorphous tips lie outside the bounds of M-D theory.
This is an important result because pulloff forces are frequently used to measure the work of adhesion.
Based on continuum theory, one would expect that the uncertainty in such measurements is less than the difference between JKR and DMT predictions.
Our results show factor of two deviations for stepped tips, which may be common in scanning probe experiments.
Other simulations showed that the stepped tip values are strongly dependent on the size of the first terrace, as well as any tilt of the terraces or incommensurability.

It might seem surprising that the stepped tip has a smaller pulloff force than the bent tip, because the entire first terrace can adhere without any elastic energy penalty.
However this effect is overcome by the greater contact size for bent tips: The radius of the first terrace, $r_t \sim 6\ \sigma$, is smaller than the values of $r_b$ and $r_c$ at pulloff for bent tips.
As the adhesion is decreased, the predicted contact size at pulloff will drop below $r_t$ and the stepped tip may then have a larger pulloff force than
predicted.
This limit can also be reached by increasing the width of the first terrace.
For a tip that lies entirely within a sphere of radius $R$, $r_t^2 < R^2-(R-d)^2\approx 2dR$ where $d$ is the layer spacing in the crystal.
For our geometry this corresponds to $r_t < 12\ \sigma$, which is about twice the value for our tip.
As noted above, terraces with even smaller radii may be preferentially selected for imaging studies and will lead to lower $|N_c|$.

As $r_{\rm cut}$ increases, it becomes harder for the M-D model to simultaneously fit both the radii and the pulloff force.
Figure \ref{fig:cutoff22} shows simulation data for a bent commensurate tip with $r_{\rm cut}=2.2\ \sigma$. 
Using the value of $h_0=1.2\ \sigma$ (Fig. \ref{fig:cutoff}) reproduces all radii fairly well at large loads, but gives a substantially different pulloff force,
$-906\ \epsilon/\sigma$ instead of $-872\ \epsilon/\sigma$.
Decreasing $h_0$ to $0.8\ \sigma$ fits the pulloff force, and improves the fit to $r_a$ at low $N$.
However, the predicted values for $r_a$ at large $N$ are slightly too high and the values for $r_c$ are shifted far below ($\sim 2-3\ \sigma$) simulation data.
This failure is not surprising given the crude approximation for adhesive interactions in M-D theory.
As might be expected, the best values for the pulloff force are obtained 
by fitting the region near the peak in the force, rather than the weak tail (Fig. \ref{fig:cutoff}).

It should be noted that for bent crystalline and amorphous tips, all of our results for $r$ can be fit accurately to M-D theory if $E^*$, $w$, and $R$ are taken as adjustable parameters.
The typical magnitude of adjustments is 10 to 30\%, which is comparable to typical experimental uncertainties for these quantities in nanoscale experiments.
Indeed one of the common goals of these experiments is to determine any scale dependence in continuum parameters.
For this reason it would be difficult to test continuum theory in scanning probe experiments even if $r$ could be measured directly.

Experiments can more easily access the variation of normal displacement with load.
This requires subtracting the height change due to the normal compliance of the machine controlling the tip, which is difficult for standard AFM cantilevers, but possible in stiffer devices \cite{jarvis93,kiely98,asif01}.
The absolute zero of $\delta$ is not fixed, but must be fitted to theory.
Figure \ref{fig:r05dofN} shows two measures of the normal displacement in our simulations.
One, $\delta_{\rm tip}$, corresponds to the experimentally accessible tip displacement.
We associate the zero of $\delta_{\rm tip}$ with the point where the first substrate atom exerts a repulsive force.
The second, $\delta_{\rm sur}$, is the actual depression of the lowest substrate atom on the surface relative to the undeformed surface.  
The two differ because of the interfacial compliance normal to the surface,
which is assumed to vanish in continuum theory.

The simulation results for $\delta$ are more sensitive to the finite sample depth $D$ and lateral periodicity $L $ than other quantities.
To account for this, the predictions of M-D and JKR theory are shifted by the leading analytic corrections in $a/D$ \cite{johnson01,sridhar04,adams06}:
\begin{equation}
\delta = \delta_{H}(1-b\frac{a}{D})+\delta_{\rm adhesion}(1-d\frac{a}{D}) \, ,
\label{eq:depth}
\end{equation}
where $\delta_H$ is the Hertz prediction (Eq. (\ref{eq:hertzd})),
$b$ and $d$ are fit parameters,
and $\delta_{\rm adhesion} =-\sqrt{2w\pi a /{E}^*} $ for JKR theory
and $-(2\sigma_0/E^*)\sqrt{c^2- a^2}$ for M-D theory.
We obtained $b=0.8$ from simulations with dense, non-adhesive tips (Fig. \ref{fig:hertz}) and $d=0.3$ from simulations with dense, adhesive tips.
Results for dense tips are then indistinguishable from the fit lines in
Figs. \ref{fig:hertz}(a) and \ref{fig:r05dofN}.
Values of $b$ and $d$ from numerical studies of continuum equations are of the same order \cite{johnson01,sridhar04,adams06}, but to our knowledge these calculations have not considered our geometry where $L\approx D$.

With the finite-size corrections, continuum results for $\delta$ lie very close to the simulation data for bent tips.
Agreement is best for $\delta_{\rm sur}$ because neither it nor continuum theory include the interfacial compliance.
The choice of zero for $\delta_{\rm tip}$ is not precisely consistent with M-D theory.
The first repulsive interaction would correspond to the first non-zero $r_a$, which occurs at a slightly negative $\delta$ in M-D theory.
For the parameters used here, the shift in $\delta$ is about 0.24 $\sigma$,
which is about twice the discrepancy between $\delta_{\rm tip}$ and M-D theory
at large loads.
This implies that the interfacial compliance produces a comparable correction.

The effects of interfacial compliance are biggest for the most negative values of $\delta_{\rm tip}$.
Here $\delta_{\rm tip}$ decreases monotonically as $N$ increases to zero.
In contrast, $\delta_{\rm sur}$ is flat and then increases back towards zero.
In this regime, $r_a =0$ and the only interaction comes from the attractive tail of the potential.
The net adhesive force gradually decreases ($N$ increases) as the magnitude of the separation between tip and substrate, $\Delta \delta \equiv \delta_{\rm tip}-\delta_{\rm sur}$, increases (inset).
Most of this change occurs in $\delta_{\rm tip}$, while the displacement of the surface relaxes back to zero as the attractive force on it goes to zero.
The JG extension to M-D theory is expressed in terms of $\delta_{\rm tip}$ and provides a good description of its change with $N$ (dashed line), even with the assumption of a constant attractive $\sigma_0$.

For amorphous tips, both values of $\delta$ are substantially above
M-D theory.
The shifts are bigger than in the non-adhesive case, about $0.35\ \sigma$
vs. 0.25$\ \sigma$.
This is correlated with the larger than predicted pulloff force.
Based on the increase in $|N_c|$, the effective work of adhesion appears to be larger by about 30\% than that measured for a flat surface.
The stronger adhesive contribution to the total normal force leads to a larger value of $\delta$.
One may understand these changes in terms of the effect of surface roughness.
Atomic-scale roughness on a nominally flat surface prevents many atoms from optimizing their binding energy.
As $\delta$ and the contact area shrink, the long wavelength height fluctuations become irrelevant and no longer prevent the few atoms remaining in the contact from adhering. 
Thus while the large load values of $\delta$ can be fit to continuum predictions with the measured $w$ and a simple shift in the origin of $\delta$, the small $N$ values correspond to a larger work of adhesion.
The magnitude of the increase ($\sim 30$\%) is modest given that the incommensurate tip has about twice as large a $w$ as the amorphous tip for the same interaction energy $\epsilon_i$.

The data for stepped tips are qualitatively different than the others.
As for the non-adhesive case, $\delta$ is lower than the continuum prediction at large loads, because the flat tip is harder to push into the substrate.
The deviation increases rapidly at negative loads, with a sharp drop near
$N=0$ where the contact shrinks to the first terrace.
As noted above, the radius of the first terrace is smaller than the radius at pulloff predicted by continuum theory, and the pulloff force is less than half the predicted value.

\subsection{Friction and lateral stiffness}
\label{sec:adherefriction}

Scanning probe microscopes can most easily detect friction forces and the lateral compliance of the tip/substrate system \cite{carpick97,carpick97b}.
Figure \ref{fig:r05FofN} shows $F$ as a function of load for five different adhesive tips.
As for non-adhesive tips, tip geometry has a much larger effect on $F$ than other quantities, and values for bent commensurate and incommensurate tips differ by two orders of magnitude.

Since the friction was measured at constant load (Sec. \ref{sec:method}), only values in the stable regime, $d\delta_{\rm tip}/dN >0 $ could be obtained.
Even in this regime, we found that the tip tended to detach at $N > N_c$.
This was particularly pronounced for commensurate tips.
Indeed the bent and stepped commensurate tips detached after the first peak in the friction force for all the negative loads shown in Fig. \ref{fig:r05FofN}(a).
At loads closer to the pulloff force, detachment occurred at even smaller lateral displacements.

As noted above, bent commensurate tips have the strongest adhesion energy because all atoms can simultaneously optimize their binding.
For the same reason, the adhesion energy changes rapidly as atoms are displaced laterally away from the optimum position, allowing pulloff above the expected
$N_c$.
The extent of the change depends on the sliding direction and registry
\cite{harrison99,muser03acp,muser04}.
We consider sliding in the (100) direction (Sec. \ref{sec:method}), where atoms move from points centered between substrate atoms towards points directly over substrate atoms.
This greatly reduces the binding energy, leading to detachment at less than half the pulloff force.
Note that changes in binding energy with lateral displacement lead directly to a lateral friction force \cite{muser03acp} and the bent and stepped commensurate tips also have the highest friction.
We suspect that tips with higher friction may generally have a tendency to detach farther above $N_c$ during sliding than other tips.

At the macro scale, friction is usually assumed to be directly proportional to load.
All the tips have substantial friction at zero load, due to adhesion.
The friction force also varies non-linearly with $N$, showing discrete jumps for the stepped tip, and curvature for the other tips, particularly near $N_c$.
The curvature at small $N$ is reminiscent of the dependence of radius on load.
Several authors have fit AFM friction data assuming that $F$ scales with contact area, and using a continuum model to determine the area as a function of load
\cite{carpick97,carpick97b,pietrement01,lantz97,carpick04,carpick04b,schwarz97,schwarz97b}.
The dotted lines in Fig. \ref{fig:r05FofN} show attempts to fit $F$ using the area from JKR theory with $w$ adjusted to fit the pulloff force and the proportionality constant chosen to fit the friction at high loads.
None of the data is well fit by this approach.
The solid lines show fits to an expression suggested by Schwarz \cite{schwarz03} that allows one to simply interpolate between DMT and JKR limits as a parameter $t_1$ is increased from 0 to 1.
Reasonable fits are possible for non-stepped tips when this extra degree of freedom is allowed.
Values of the friction per unit area are within 25\% of those determined below from the true area (Table II),
but the values of $w$ and $t_1$ do not correspond to any real microscopic parameters.
For example, for the three non-stepped tips where a direct measurement gives $w=0.46\ \epsilon/\sigma^2$, the fits give $w=0.41\ \epsilon/\sigma^2$, $0.55\ \epsilon/\sigma^2$ and $0.94\ \epsilon/\sigma^2$ for commensurate, amorphous and incommensurate tips, respectively.
The value of $t_1$ varies from 0.1 to 0.5 to 1.5, respectively.
Note that the value of $t_1=1.5$ for incommensurate tips is outside the physical range.
In this case, and some experiments \cite{carpick04}, the friction rises more slowly than the JKR prediction, while M-D theory and simpler interpolation schemes \cite{schwarz03,carpick99} always give a steeper rise.
Such data seem inconsistent with $F$ scaling with area.

Our simulations allow us to test the relationship between $F$ and area without any fit parameters.
However, it is not obvious which radius should be used to determine area.
Figure \ref{fig:r05Fofr} shows friction plotted against $r_a^2$, $r_b^2$ and
$r_c^2$.
The stepped tip is not shown, since only $r_c$ is easily defined and it
increases in one discontinuous jump over the range studied.
For all other tips the friction is remarkably linear when plotted
against any choice of radius squared.
In contrast, plots of $N$ vs. $r^2$ show significant curvature.

For $F$ to be proportional to area, the curves should also pass through the origin.
This condition is most closely met by $r_a^2$, except for the incommensurate case.
The idea that friction should be proportional to the area where atoms are pushed into repulsion seems natural.
The other radii include attractive regions where the surfaces may separate far enough that the variation of force with lateral displacement is greatly reduced or may even change phase.
However, in some cases the extrapolated friction remains finite as $r_a$ goes to zero and in others it appears to vanish at finite $r_a$.

Also shown in Fig. \ref{fig:r05Fofr} are results for non-adhesive tips (open circles).
Values for incommensurate and amorphous tips were doubled to make the linearity of the data more apparent.
Only the bent commensurate data shows significant curvature, primarily at small $N$.
As noted in Sec. \ref{sec:nonadhere}, friction is proportional to load for this tip, and the curvature is consistent with this and $a^2 \sim N^{2/3}$.
No curvature is visible when adhesion is added, but the fact that the linear fit extrapolates to $F=0$ at $r_a/\sigma \sim 4$ suggests that the linearity might break down if data could be obtained at lower loads.

Results for adhesive and non-adhesive amorphous tips can be fit by lines through the origin (not shown), within numerical uncertainty.
The same applies for non-adhesive bent incommensurate tips, but adding adhesion shifts the intercept to a positive force at $r_a=0$.
It would be interesting to determine whether this extrapolation is valid.
Friction can be observed with purely adhesive interactions, since the only requirement is that the magnitude of the energy varies with lateral displacement \cite{muser03acp}.
However, it may also be that the friction on incommensurate tips curves rapidly to zero as $r_a \rightarrow 0$.
Indeed one might expect that for very small contacts the tip should behave more like a commensurate tip, leading to a more rapid change of $F$ with area or load.
The only way to access smaller $r_a$ is to control the tip height instead of the normal load.
This is known to affect the measured friction force \cite{muser04}, and most experiments are not stiff enough to access this regime.

Figure \ref{fig:r05kofr} shows the lateral stiffness as a function of the three characteristic radii.
Except for the stepped tip (not shown), $k$ rises linearly with each of the radii.
As for non-adhesive tips, the slope is much smaller than the value $8G^*$ predicted by continuum theory (solid lines) because of the interfacial compliance (Eq. (\ref{eq:stiff})).
The intercept is also generally different from the origin, although it comes closest to the origin for $r_a$.
As for friction, it seems that the repulsive regions produce the dominant
contribution to the stiffness.

Results for non-adhesive tips are also included in Fig. \ref{fig:r05kofr}.
They also are linear over the whole range, and the fits reach $k=0$ at a finite radius $a_k$.
This lack of any stiffness between contacting surfaces seems surprising.
Note however that linear fits to Fig. \ref{fig:hertz}(b) also would suggest that
the radius approached a non-zero value, $a_0$, in the limit of zero load.
Moreover, the values of $a_0$ and $a_k$ follow the same trends with tip geometry and have similar sizes.

The finite values of $a_0$ and $a_k$ can be understood from the finite range
of the repulsive part of the interaction.
As long as atoms are separated by less than $r_{\rm cut}$ there is a finite interaction and the atoms are considered inside $a$.
However, the force falls rapidly with separation, and atoms near this outer limit contribute little to the friction and stiffness.
If $\delta h$ is the distance the separation must decrease to get a significant interaction, then $a_0 = (2R\delta h)^{1/2}$ at the point where the first significant force is felt.
Taking the estimate of $\delta h=0.04\ \sigma$ from Sec. \ref{sec:nonadhere},
then $a_0 \sim 3\ \sigma$, which is comparable to the observed
values of $a_0$ and $a_k$.
The shift is smaller for the adhesive case because there are still
strong interactions when $r_a$ goes to zero.
The larger shifts for the amorphous tip may reflect roughness, since the first point to contact need not be at the origin.
We conclude that the linear fits in Fig. \ref{fig:r05kofr} go to zero at finite radius, because $r_a$ overestimates the size of the region that makes significant contributions to forces, particularly for non-adhesive tips.
Note that the plots for friction with non-adhesive tips (Fig. \ref{fig:r05Fofr}) are also consistent with an offset, but that the offset appears much smaller when plotted as radius squared.

The slope of the curves in Fig. \ref{fig:r05Fofr} can be used to define a differential friction force per unit area or yield stress $\tau_f \equiv \partial F/ \partial \pi r_a^2$.
It is interesting to compare the magnitude of these values (Table II) to the bulk yield stress of the substrate $\tau_y$.
Assuming Lennard-Jones interactions, the ideal yield stress of an fcc crystal in the same shearing direction is 4 to 10 $\epsilon \sigma^{3}$, depending on whether the normal load or volume is held fixed.
The commensurate tip is closest to a continuation of the sample, and the force on all atoms adds coherently.
As a result $\tau_f$ is of the same order as $\tau_y$, even for the non-adhesive tip.
Values for adhesive amorphous and incommensurate tips are about one and two orders of magnitude smaller than $\tau_y$, respectively.
This reflects the fact that the tip atoms can not optimize their registry with the substrate.
Removing adhesive interactions reduces $\tau_y$ by an additional factor of about four in both cases.

In continuum theory, $k=8G^* r$, and the slope of fits in Fig. \ref{fig:r05kofr} could then be used to determine the effective shear modulus $G^*$.
However, as noted above, the interfacial compliance leads to much lower stiffnesses.
To illustrate the magnitude of the change we quote values of
$G' \equiv (1/8) \partial k /\partial r$ in  Table II.
All values are below the true shear modulus $G^* =18.3\ \epsilon/\sigma^3$ obtained from the substrate compliance alone (Fig. \ref{fig:hertz}).
As always, results for bent commensurate tips come closest to continuum theory
with $G'/G^* \sim 0.7$.
Values for adhesive amorphous and incommensurate tips are depressed by factors of 3 and 20 respectively, and removing adhesion suppresses the value for amorphous tips by another factor of four.

Carpick et al. noted that if friction scales with area and $k$ with radius, then the ratio $F/k^2$ should be constant \cite{carpick97thesis,carpick97b,carpick04,carpick04b}.
Defining the frictional force per unit area as $\tau_f$ and using the expression for $k$ from continuum theory, one finds $64 {G^*}^2 F/\pi k^2 = \tau_f$. 
In principal, this allows continuum predictions to be checked and $\tau_f$ to be determined without direct measurement of contact size.
Figure \ref{fig:r05ratofN} shows:
\begin{equation}
\tau_f^{\rm eff} \equiv 64 (G^*)^2 F/\pi k^2
\label{eq:taueff}
\end{equation}
as a function of $N$ for different tip geometries and interactions.
Except for the non-adhesive stepped tip,
the value of $F /k^2$ is fairly constant at large loads,
within our numerical accuracy.
Some of the curves rise at small loads because the radius at which $F$
reaches zero in Fig. \ref{fig:r05Fofr} tends to be smaller than that where
$k$ reaches zero in Fig. \ref{fig:r05kofr}.
These small radii are where continuum theory would be expected to be least accurate.
Note that the deviations are larger for the non-adhesive tips, perhaps because the data extends to smaller radii.

The data for stepped tips are of particular interest because the contact radius jumps in one discrete step from the radius of the first terrace to the radius of the second.
The friction and stiffness also show discontinuous jumps.
Nonetheless, the ratio $F/k^2$ varies rather smoothly and even has
numerical values close to those for other tips.
The most noticeable difference is that the data for the nonadhesive stepped tip
rises linearly with load, while all other tips tend to a constant
at high load.
These results clearly demonstrate that success at fitting derived quantities like $F$ and $k$ need not imply that the true contact area is following continuum theory.

The curves for $\tau_f^{\rm eff}$ in Fig. \ref{fig:r05ratofN} are all much higher than values of the frictional stress $\tau_f$ obtained directly from the friction and area (Table II).
Even the trends with tip structure are different.
The directly measured frictional stress decreases from bent commensurate
to amorphous to bent incommensurate, while $\tau_f^{\rm eff}$ is
largest for the amorphous and smallest for the bent commensurate tip.
These deviations from the continuum relation
are directly related to the interfacial compliance $k_{\rm i}$.
The continuum expression for the lateral stiffness neglects $k_{\rm i}$ and gives too small a radius at each load.
This in turn over-estimates the frictional stress by up to two orders of magnitude.
Similar effects are likely to occur in experimental data.

Experimental plots of $F/k^2$ have been obtained for silicon-nitride tips on mica and sodium chloride \cite{carpick97thesis,carpick04,carpick04b} and on carbon fibers \cite{pietrement01}.
Data for carbon fibers and mica in air showed a rapid rise with decreasing $N$ at low loads \cite{pietrement01,carpick97thesis}.
For mica the increase is almost an order of magnitude, which is comparable to our results for non-adhesive bent incommensurate tips.
This correspondence may seem surprising given that the experiments measured adhesion in air.
However, the adhesive force was mainly from long-range capillary forces that operate outside the area of contact.
Following DMT theory, they can be treated as a simple additive load that does not affect the contact geometry.
In contrast, data for mica in vacuum is well fit by JKR theory, implying a strong adhesion within the contact \cite{carpick04,carpick04b}.
The measured value of $F/k^2$ is nearly constant for this system, just as in our results for most adhesive tips.
Results for carbon fibers in vacuum \cite{pietrement01} show a linear rise like that seen for nonadhesive stepped tips.

From a continuum analysis of the carbon fiber data, the frictional
stress was estimated to be
$\tau_f \sim 300$ MPa assuming a bulk shear stress of $G^*=9.5$ GPa
\cite{pietrement01}.
Note that Fig. \ref{fig:r05ratofN} would suggest $\tau_f/G^* \sim 0.1$ to 0.3, while the true values (Table II) are as low as $0.0002$.
The data on carbon fibers could be fit with the bulk shear modulus, but data on mica and NaCl \cite{carpick97thesis,carpick04} indicated that $G^*$ was 3 to 6 times smaller than bulk values.
Our results show that the interfacial compliance can easily lead to reductions of this magnitude and a corresponding increase in $\tau_f$,
and that care must be taken in interpreting experiments with continuum models.

\section{Discussion and Conclusion}
\label{sec:conclusions}

\begin{figure}
\begin{center}
\includegraphics[width=14cm]{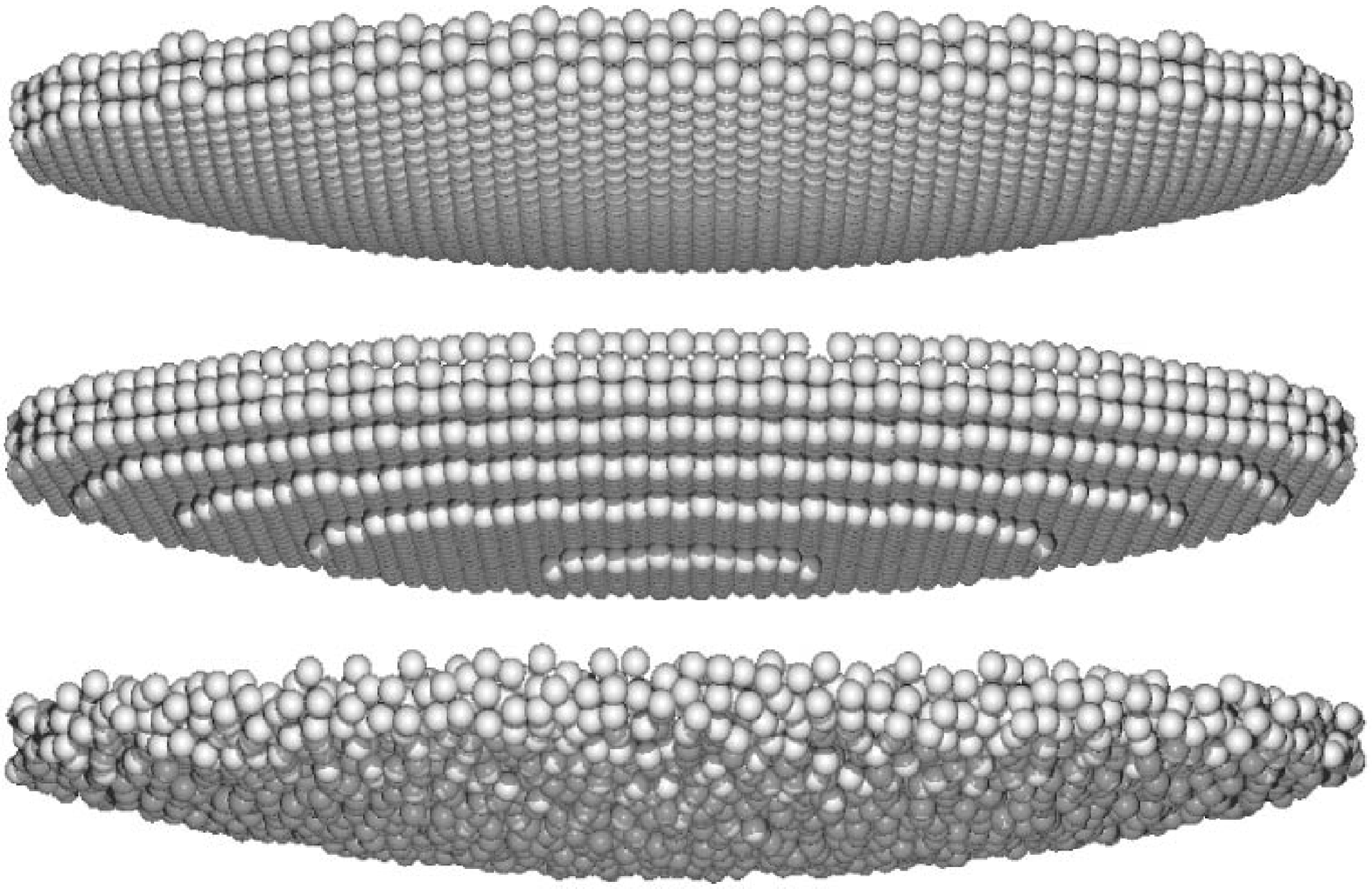}
\caption{
Snapshots of atoms near the center regions (diameter 50$\ \sigma$) of spherical tips with average radius $R$=100$\ \sigma$. From top to bottom, tips are made by bending a crystal, cutting a crystal or cutting an amorphous solid.
Three ratios $\eta$ of the atomic spacing in bent crystals to that in the
substrate are considered; a dense case $\eta=0.05$,
a commensurate case $\eta=1$, and an incommensurate case $\eta=0.94437$.
The step structure of cut crystalline tips is not unique, leading to
variations in their behavior.
}
\label{fig:tips}
\end{center}
\end{figure}

The results described above show that many different effects can lead to deviations between atomistic behavior and continuum theory and
quantify how they depend on tip geometry for simple interaction
potentials (Fig. \ref{fig:tips}).
In general, the smallest deviations are observed for the idealized model of a dense tip whose atoms form a nearly continuous sphere, although this tip has nearly zero friction and lateral stiffness.
Deviations increase as the geometry is varied from a bent commensurate to a bent incommensurate to an amorphous tip, and stepped tips exhibit qualitatively different behavior.
Tip geometry has the smallest effect on the normal displacement and normal stiffness (Fig. \ref{fig:hertz} and \ref{fig:r05dofN}) because they reflect an average response of the entire contact.
Friction and lateral stiffness are most affected (Fig. \ref{fig:hertz} and \ref{fig:r05FofN}), because they depend on the detailed lateral interlocking of atoms at the interface.

\begin{figure}
\begin{center}
\includegraphics[width=14cm]{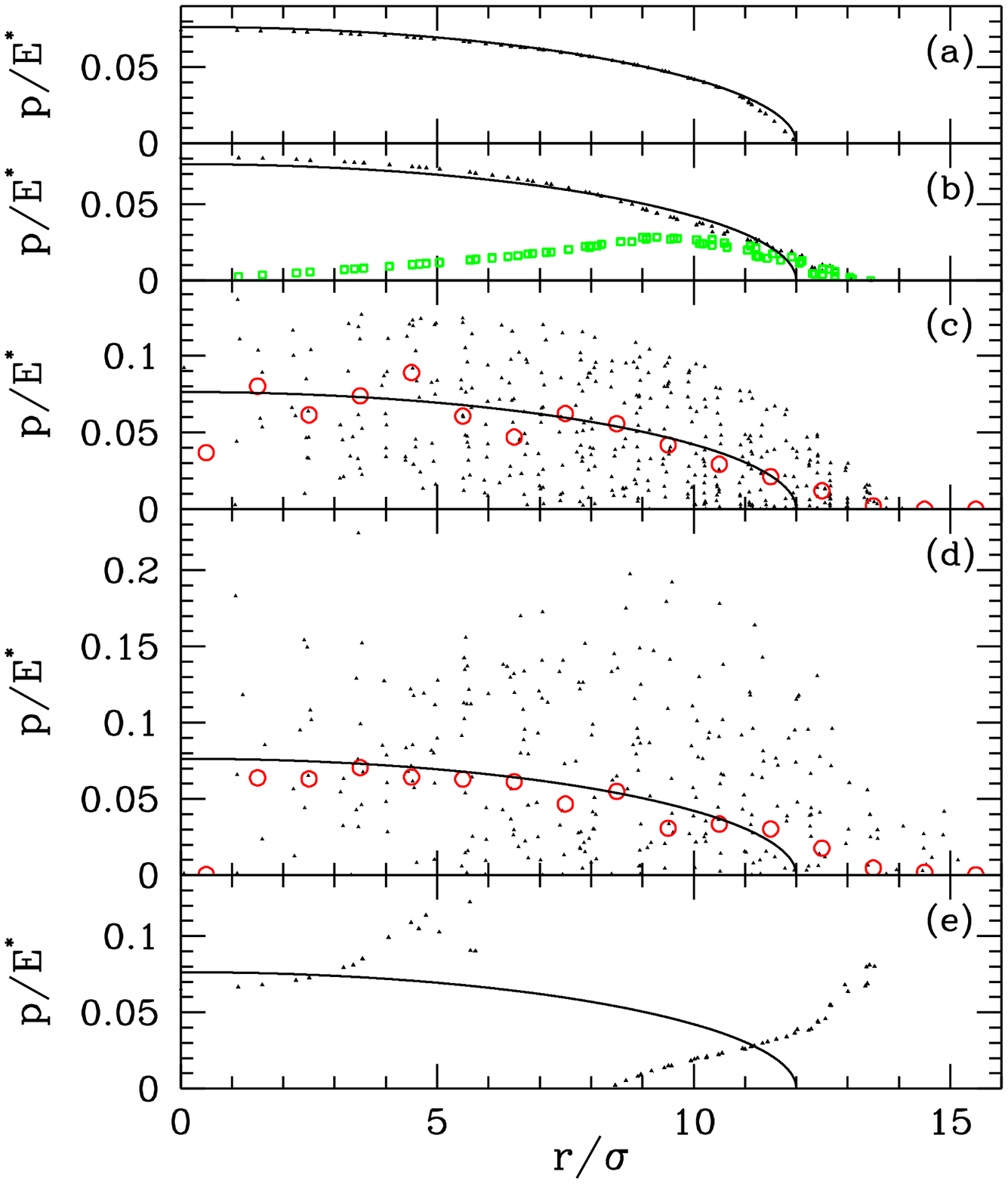}
\caption{
(Color online)
Local normal pressure vs. radius for five different tip geometries (a) dense tip, (b) bent commensurate crystal, (c) bent incommensurate crystal, (d) amorphous, and (e) stepped crystal.
All are non-adhesive, and have the same nominal radius $R=100\ \sigma$, $\epsilon_i/\epsilon=1$, and normalized load $N/(R^2 E^*)=0.0018$.
Solid lines show the prediction of Hertz theory, dots show the pressure on each surface atom, and circles in (c) and (d) show the mean pressure in radial bins of width $\sigma$.
Squares in (b) show the component of the tangential force directed radially
from the center of the contact.  The azimuthal component is nearly zero,
and tangential forces are much smaller for other tips.
}
\label{fig:hertzpofr}
\end{center}
\end{figure}

\begin{figure}
\begin{center}
\includegraphics[width=12cm]{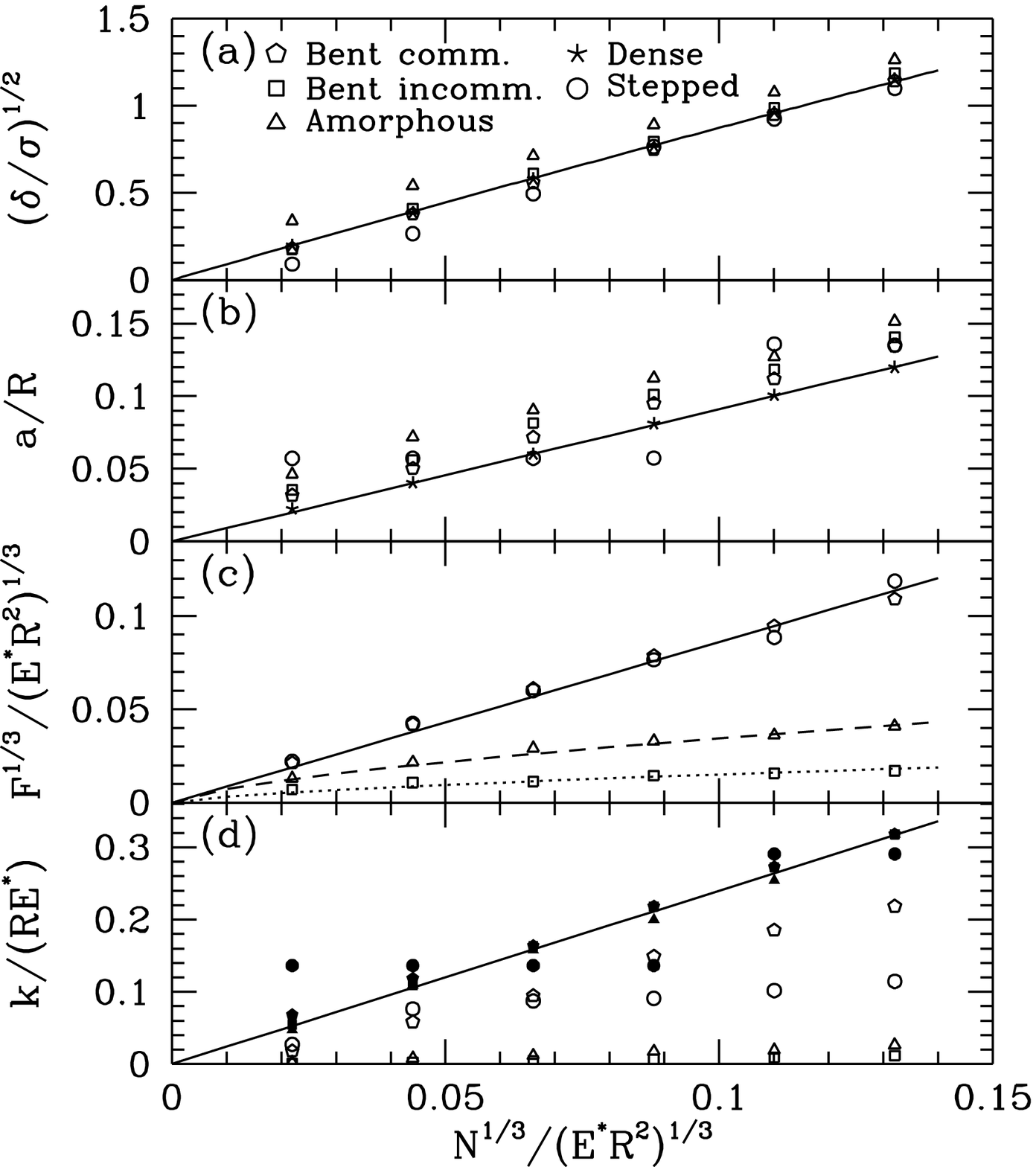}
\caption{
Dimensionless plots of powers of (a) normal displacement $\delta$,
(b) contact radius $a$, (c) static friction $F$, and
(d) lateral stiffness $k$ vs. the cube root of the normal force $N$ for
non-adhesive tips with $R=100\ \sigma$.
The powers of $\delta$, $a$ and $k$ are chosen so that
continuum theory
predicts the indicated straight solid lines.
In (c), the solid line corresponds to $F\propto N$, while
dashed and dotted lines are fits to $F \propto N^{2/3}$.
The continuum predictions for $\delta$ and $r$ are followed by a dense 
tip (stars).
Also shown are results for bent commensurate (pentagons),
bent incommensurate (squares), amorphous (triangles)
and stepped commensurate tips (circles).
In (d),
the total lateral stiffness (open symbols) lies well below the continuum
prediction because of the interfacial compliance.
The stiffness from the substrate alone (filled symbols)
scales with radius, as expected from Hertz theory.
Numerical uncertainties are comparable to the symbol size.
}
\label{fig:hertz}
\end{center}
\end{figure}

\begin{figure}
\begin{center}
\includegraphics[width=14cm]{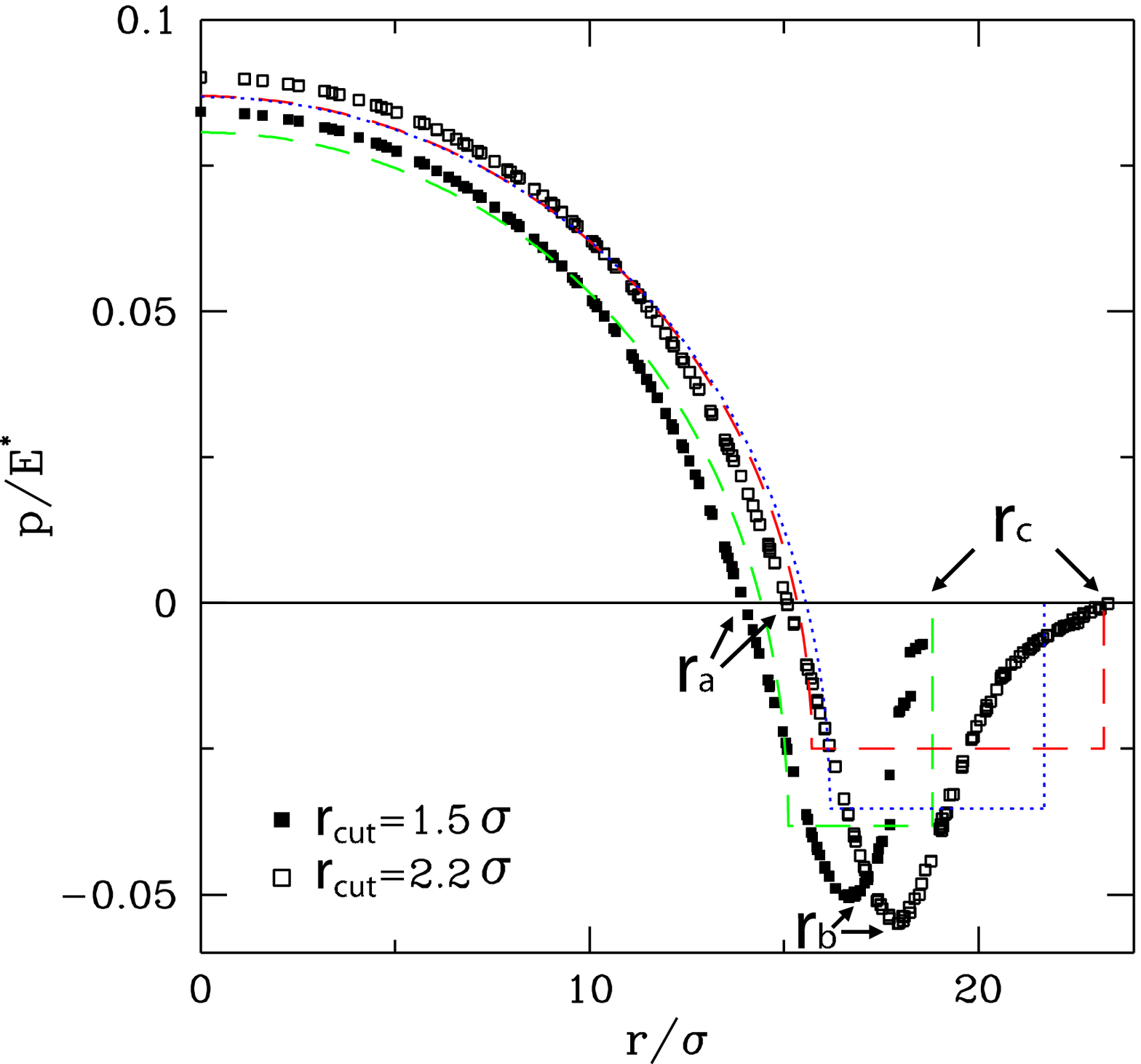}
\caption{
(Color online)
Local pressure vs. radius for bent commensurate tips with $r_{\rm cut} = 1.5\ \sigma$ (filled squares) and $r_{\rm cut}=2.2\ \sigma$ (open squares) and with $R=100\ \sigma$, $\epsilon_i/\epsilon = 0.5 $ and $N/(R^2 E^*)=0.0016$.
Dashed lines show Maugis-Dugdale fits with the measured work of adhesion and $h_0$ fit to the range of the interactions.
A dotted line shows a fit for $r_{\rm cut}=2.2\ \sigma$ with $h_0=0.8\ \sigma$ that improves agreement with the measured pulloff force (Fig. \ref{fig:cutoff22}).
Characteristic radii $r_a$, $r_b$ and $r_c$ are defined by the locations
where the radially averaged pressure first becomes zero,
is most negative, and finally vanishes, respectively.
}
\label{fig:cutoff}
\end{center}
\end{figure}

One difference between simulations and continuum theory is that the interface has a finite normal compliance.
Any realistic interactions lead to a gradual increase in repulsion with separation rather than an idealized hard-wall interaction.
In our simulations the effective range over which interactions increase is only about 4\% of the atomic spacing, yet it impacts results in several ways.
For bent commensurate tips it leads to an increase in pressure in the center of the contact (Figs. \ref{fig:hertzpofr} and \ref{fig:cutoff}).
The pressure at the edge of nonadhesive contacts drops linearly over about $2\ \sigma$, while continuum theory predicts a diverging slope.
The width of this smearing grows as the square root of the tip radius and leads to qualitative changes in the probability distribution of local pressures \cite{persson01,hyun04,pei05,luan05mrs}.
These effects could be studied in continuum theories with soft-wall interactions.
The normal interfacial compliance also leads to the offset in linear fits of $F$ vs. $r_a^2$ and $k$ vs. $r_a$ (Figs. \ref{fig:r05Fofr} and \ref{fig:r05kofr}).
Fits to the friction and local stiffness extrapolate to zero at finite values of $r_a$ because atoms at the outer edge of the repulsive range contribute to $r_a$ but interact too weakly to contribute substantially to $F$ and $k$.
This effect is largest for non-adhesive tips.

\begin{figure}
\begin{center}
\includegraphics[width=14cm]{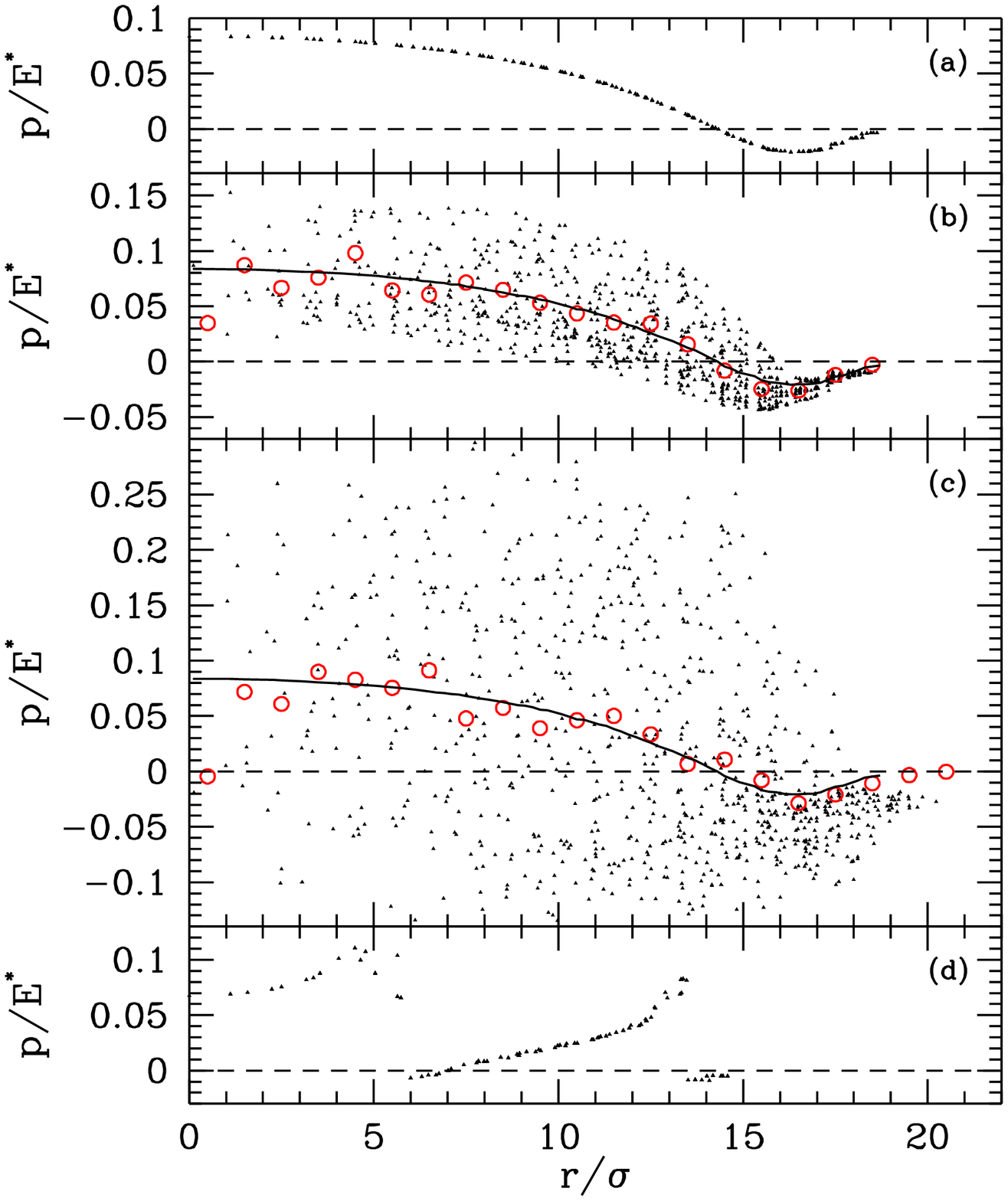}
\caption{
(Color online)
Normal pressure vs. radius for four different tip geometries:
(a) bent commensurate crystal, (b) bent incommensurate crystal,
(c) amorphous, and (d) stepped crystal.
In all cases the tip radius $R=100\ \sigma$, the normalized load $N/(R^2 E^*)=0.0023$
and the surface energy $w = 0.46\ \epsilon/\sigma^2$. 
Dots show the normal pressure on
each surface atom.
In (b) and (c), circles show the mean pressure in
radial bins of width $\sigma$, and lines show the bent commensurate
results from (a).
%Characteristic radii $r_a$, $r_b$ and $r_c$ are defined by the location
%where the radially averaged pressure first becomes zero,
%is most negative, and finally vanishes, respectively
%(arrows in panel (a)).
Horizontal dashed lines are at zero pressure.
}
\label{fig:r05pofr}
\end{center}
\end{figure}

Approximating a spherical surface by discrete atoms necessarily introduces some surface roughness.
Even bent crystalline tips have atomic scale corrugations, reflecting the variation in interaction as tip atoms move from sites nestled between substrate atoms to sites directly above.
Amorphous and stepped tips have longer wavelength roughness associated with their random or layered structures respectively.
This longer wavelength roughness has a greater effect on the contacts.
For non-adhesive interactions, incommensurate and amorphous tips have a lower central pressure and wider contact radius than predicted for ideal spheres.
These changes are qualitatively consistent with continuum calculations for spheres with random surface roughness \cite{johnson85}.
However the effective magnitude of the rms roughness $\Delta$ is smaller than expected from the atomic positions.
The correlated deviations from a sphere on stepped tips, lead to qualitative changes in the pressure distribution on the surface (Fig. \ref{fig:hertzpofr} and \ref{fig:r05pofr}).
However, these changes are also qualitatively consistent with what continuum mechanics would predict for the true tip geometry, which is closer to a flat punch than a sphere.
We conclude that the usual approximation of characterizing tips by a single spherical radius is likely to lead to substantial errors in calculated properties.
Including the true tip geometry in continuum calculations would improve their ability to describe nanometer scale behavior.
Unfortunately this is rarely done, and the atomic-scale tip geometry is rarely measured.
Recent studies of larger tips and larger scale roughness are an interesting
step in this direction \cite{thoreson06}.

Roughness also has a strong influence on the work of adhesion $w$ (Table I).
Values of $w$ were determined independently from interactions between nominally flat surfaces.
For a given interaction strength, commensurate surfaces have the highest $w$, because
each atom can optimize its binding simultaneously.
The mismatch of lattice constants in incommensurate geometries lowers $w$ by a factor of two, and an additional factor of two drop is caused by the small ($\Delta \sim 0.3\ \sigma$) height fluctuations on amorphous surfaces.
In continuum theory, these changes in $w$ should produce nearly proportional changes in pulloff force $N_c$, and tips with the same $w$ and $h_0$ should have the same $N_c$.
Measured values of $N_c$ differ from these predictions by up to a factor of two.
It is particularly significant that the dimensionless pulloff force for amorphous and stepped tips lies outside the limits provided by JKR and DMT theory.
Experimentalists often assume that these bounds place tight limits on errors in inferred values of $w$.

In the case of amorphous tips the magnitude of $N_c$ is 30\% higher than expected.
The higher than expected adhesion in small contacts may reflect a decrease in effective roughness because long-wavelength height fluctuations are suppressed.
Stepped tips show even larger deviations from continuum theory that are strongly dependent on the size of the first terraces \cite{foot1}.
Tips selected for imaging are likely to have the smallest terraces and the largest deviations from continuum theory.

\begin{figure}
\begin{center}
\includegraphics[width=14cm]{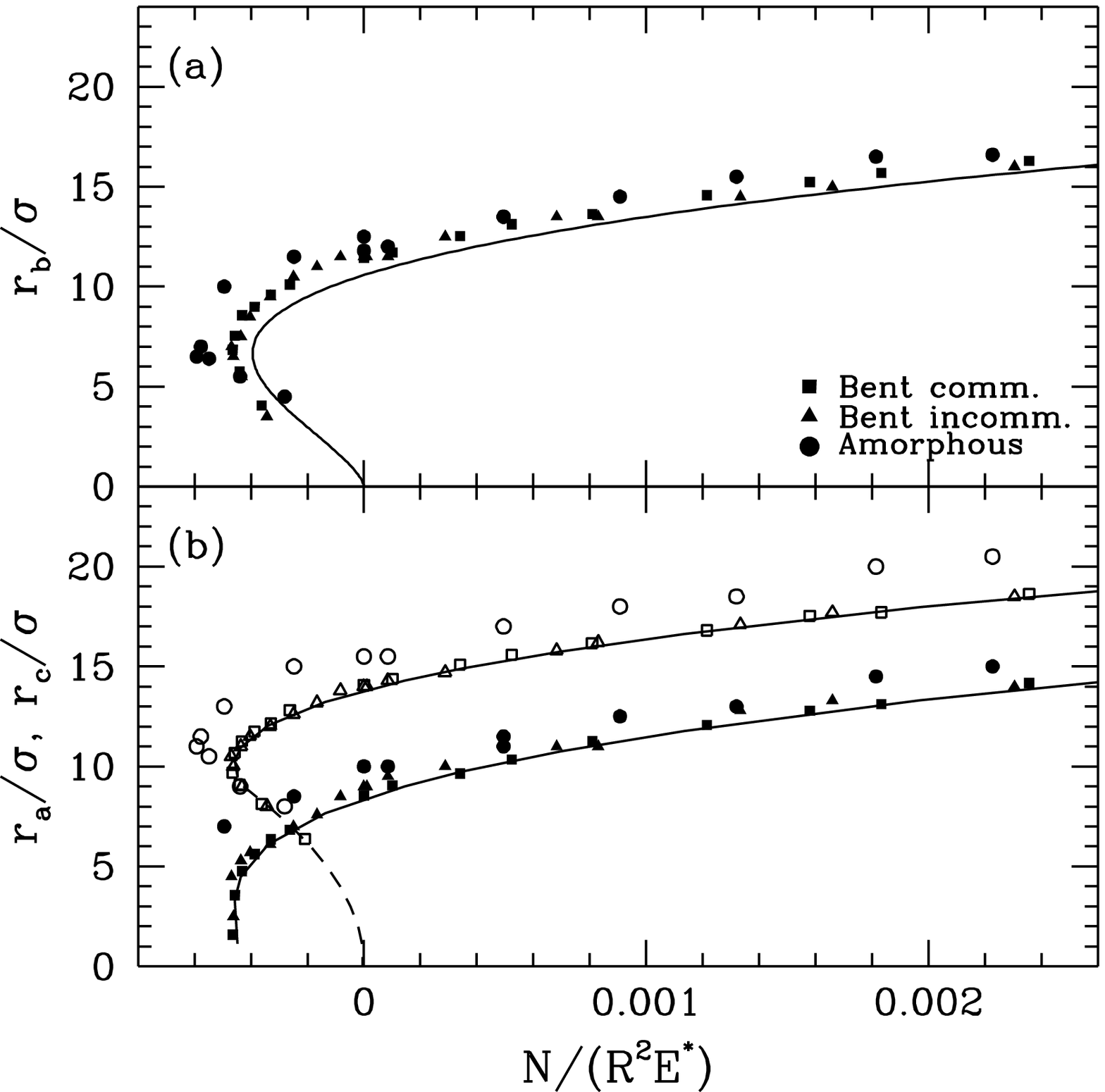}
\caption{
Variation of contact radii with external load for bent commensurate (squares),
bent incommensurate (triangles) and amorphous (circles) tips with
$R =100\ \sigma$.
In (a), the radius $r_b$ where the force is most attractive is compared to
JKR theory (solid line). 
In (b), values of $r_a$ (filled symbols) and $r_c$ (open symbols) are
compared to M-D theory (solid lines) with the JG extension for $r_c$ when $r_a=0$ (dashed line).
All continuum fits use the independently measured surface energy $w = 0.46\ \epsilon/\sigma^2$ and an interaction range $h_0 =0.5\ \sigma$ that is consistent with the potential range.
Numerical uncertainty in radii is comparable to the symbol size.
}
\label{fig:r05rofN}
\end{center}
\end{figure}

\begin{figure}
\begin{center}
\includegraphics[width=14cm]{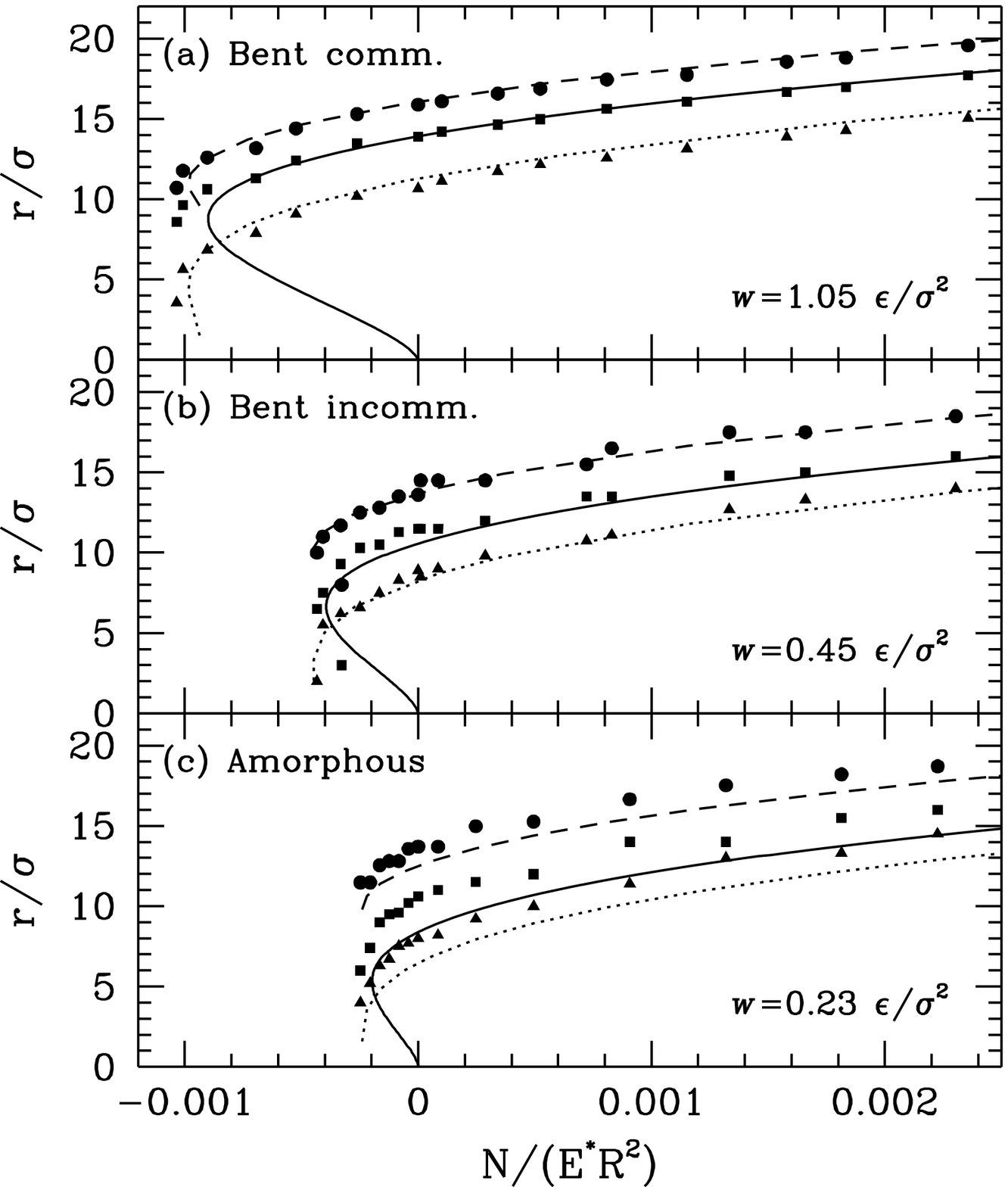}
\caption{
Contact radii as a function of normal load for (a) bent commensurate,
(b) bent incommensurate, and (c) amorphous tips with $\epsilon_i =0.5 \epsilon$.
While all have the same interaction strength, the work of adhesion
varies by more than a factor of four.
Broken lines show M-D predictions for $r_a$ (triangles) and $r_c$ (circles)
with the indicated values of $w$ and $h_0=0.5\ \sigma$.
The value of $r_b$ (squares) is closest to the JKR result (solid line).
Deviations from the continuum predictions show the same trends as in Fig. \ref{fig:r05rofN}.
Numerical uncertainty in radii is comparable to the symbol size.
}
\label{fig:e05rofN}
\end{center}
\end{figure}

\begin{figure}
\begin{center}
\includegraphics[width=14cm]{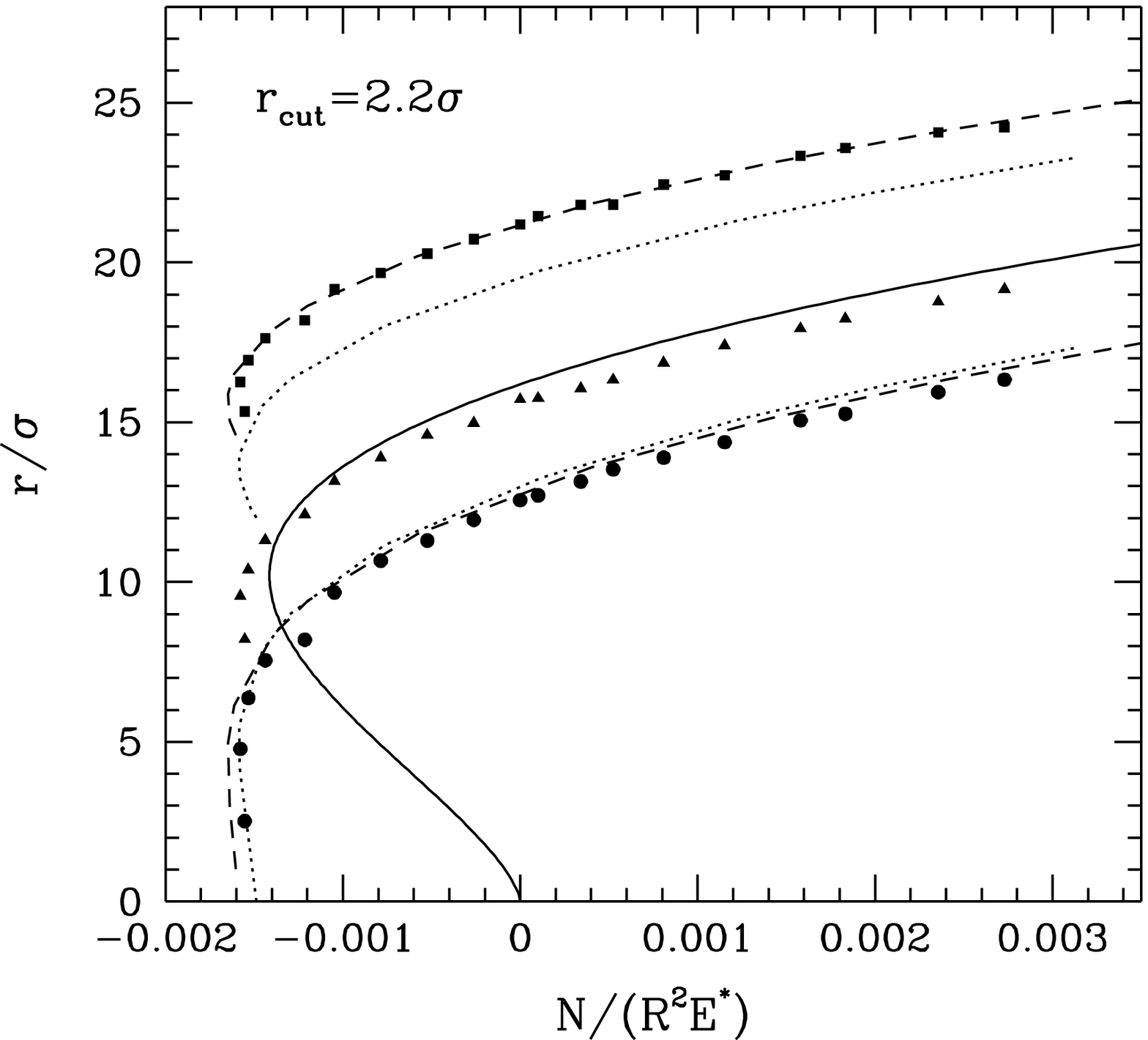}
\caption{
Fits of M-D theory (broken lines) to $r_c$ (squares) and $r_a$ (circles), and of JKR theory (solid line) to $r_b$ (triangles) for a bent commensurate tip with $r_{\rm cut}=2.2\ \sigma$.
For $h_0 =1.2\ \sigma$ (dashed line)
M-D theory fits both $r_c$ and $r_a$ at large loads, but the magnitude of the pulloff force is too large.
Decreasing $h_0$ to 0.8$\ \sigma$ (dotted lines) gives excellent agreement with the pulloff force, but not the radii (Fig. \ref{fig:cutoff}).
Numerical uncertainty in radii is comparable to the symbol size.
}
\label{fig:cutoff22}
\end{center}
\end{figure}

\begin{figure}
\begin{center}
\includegraphics[width=14cm]{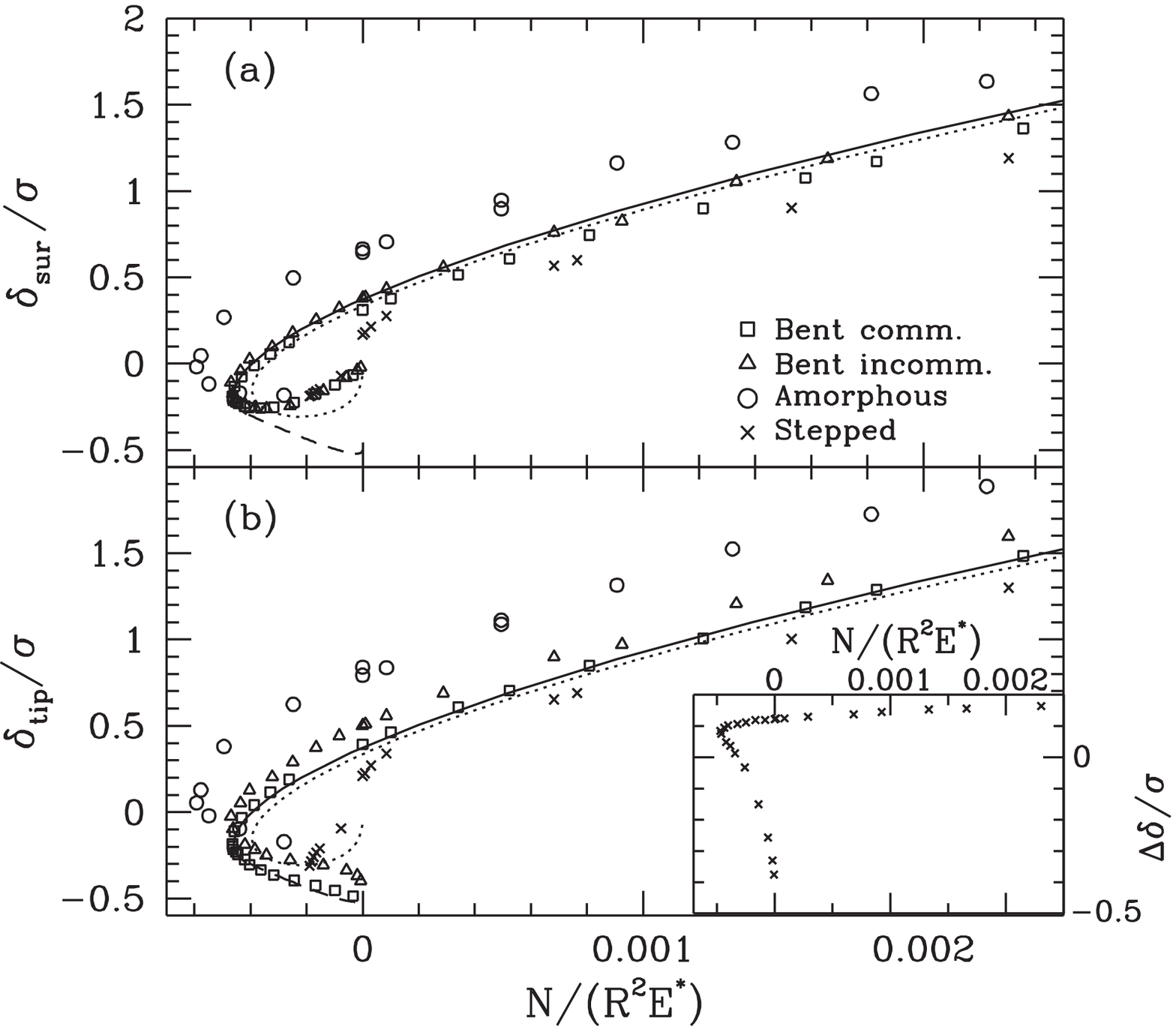}
\caption{
Normal displacement as a function of external load measured from (a) the depression of the lowest substrate atom $\delta_{\rm sur}$
and (b) the displacement of the tip relative to the height where the first substrate atom exerts a repulsive force $\delta_{\rm tip}$.
The tips are the same as in Fig. \ref{fig:r05rofN}.
The JKR prediction is indicated by dotted lines, the M-D prediction by solid
lines, and the JG extension for $r_a=0$ with dashed lines.
All are corrected for the finite dimensions of the substrate as
described in the text, and uncertainties are comparable to the symbol size.
The difference $\Delta \delta \equiv \delta_{\rm tip}-\delta_{\rm sur}$ reflects the interfacial compliance, and is shown for the bent incommensurate tip in the inset.
}
\label{fig:r05dofN}
\end{center}
\end{figure}

\begin{figure}
\begin{center}
\includegraphics[width=14cm]{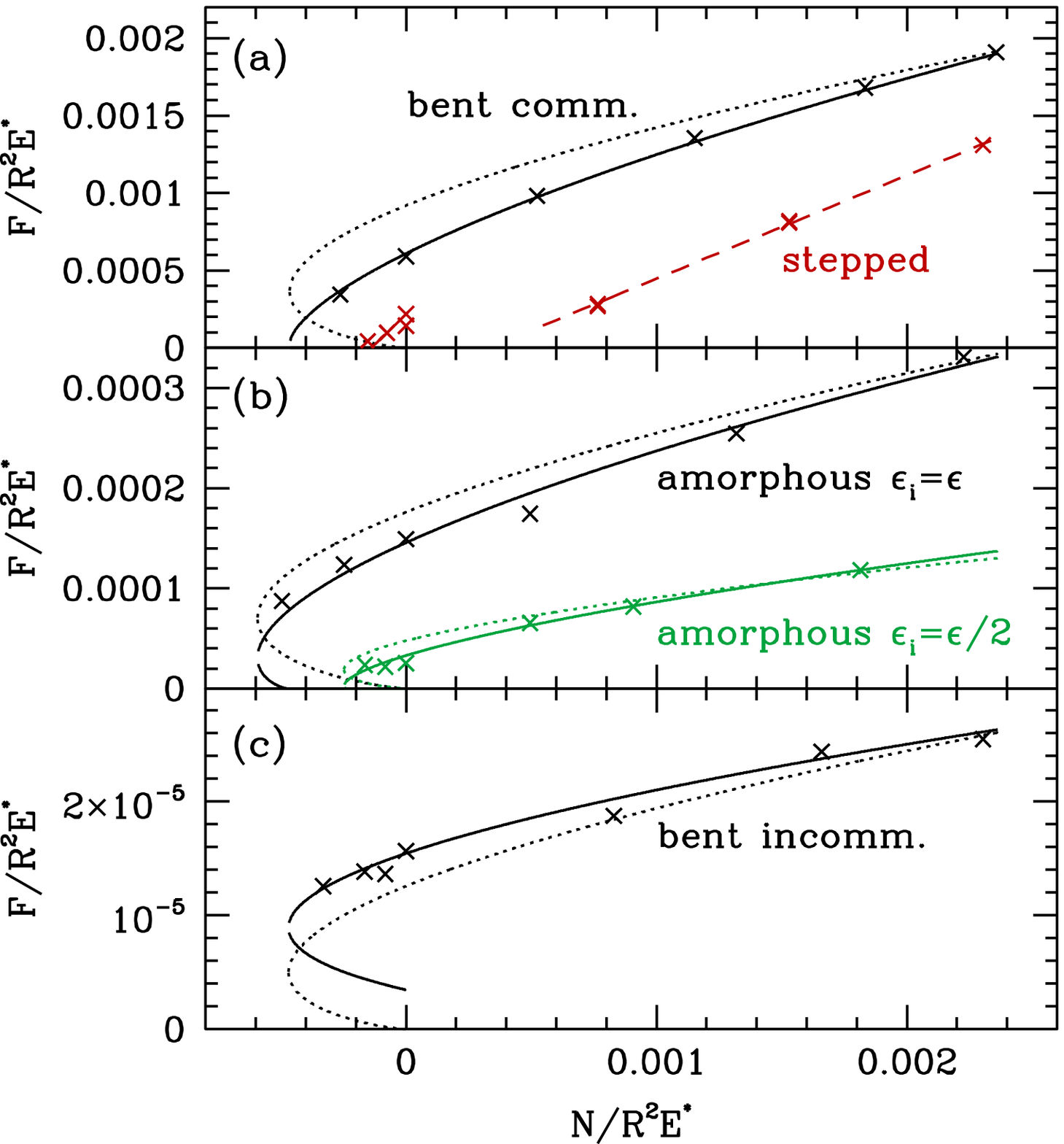}
\caption{
(Color online)
Static friction $F$ vs. load $N$ for the indicated tip geometries with same systems as in Fig. \ref{fig:r05rofN}.
Attempts to fit the data by assuming $F$ is proportional to the area predicted by JKR theory and the interpolation scheme of Ref. \cite{schwarz03}
are shown by dotted and solid lines, respectively.
Dashed lines are separate linear fits for the stepped tip for the cases where
one or two terraces are in repulsive contact.
Numerical uncertainties are comparable to the symbol size.
}
\label{fig:r05FofN}
\end{center}
\end{figure}

\begin{figure}
\begin{center}
\includegraphics[width=14cm]{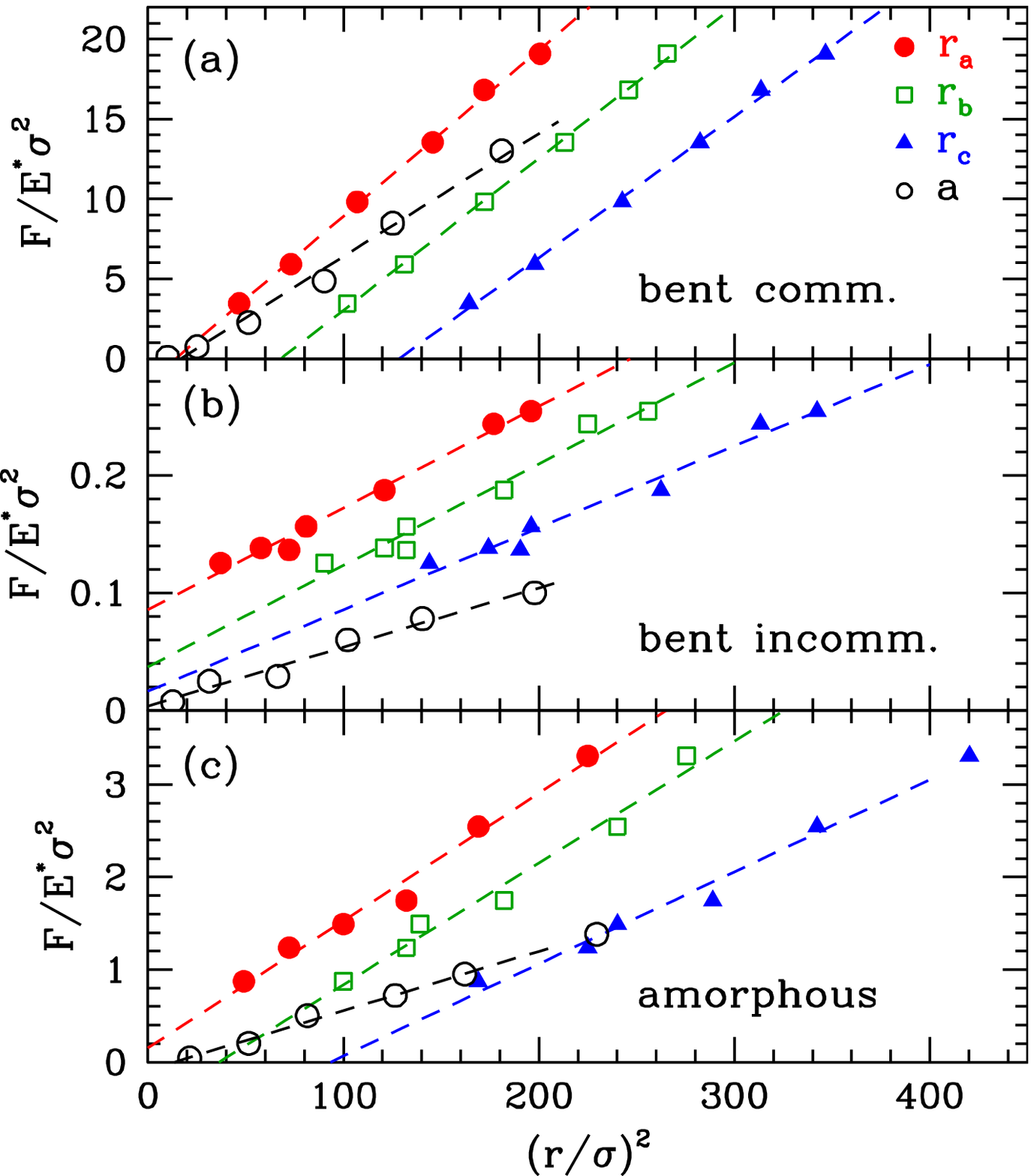}
\caption{
(Color online)
Static friction $F$ plotted against radius squared for the indicated tip geometries with same parameters as in Fig. \ref{fig:r05rofN}.
Values of $r_a^2$ (filled circles), $r_b^2$ (open squares), and $r_c^2$ (filled triangles) are shown for adhesive tips.
Open circles show $F$ vs. $a^2$ for non-adhesive tips with data in (b) and (c) multiplied by a factor of two for clarity. 
Dashed lines show unconstrained linear fits to each data set.
Numerical uncertainties are comparable to the symbol size.
}
\label{fig:r05Fofr}
\end{center}
\end{figure}

\begin{figure}
\begin{center}
\includegraphics[width=14cm]{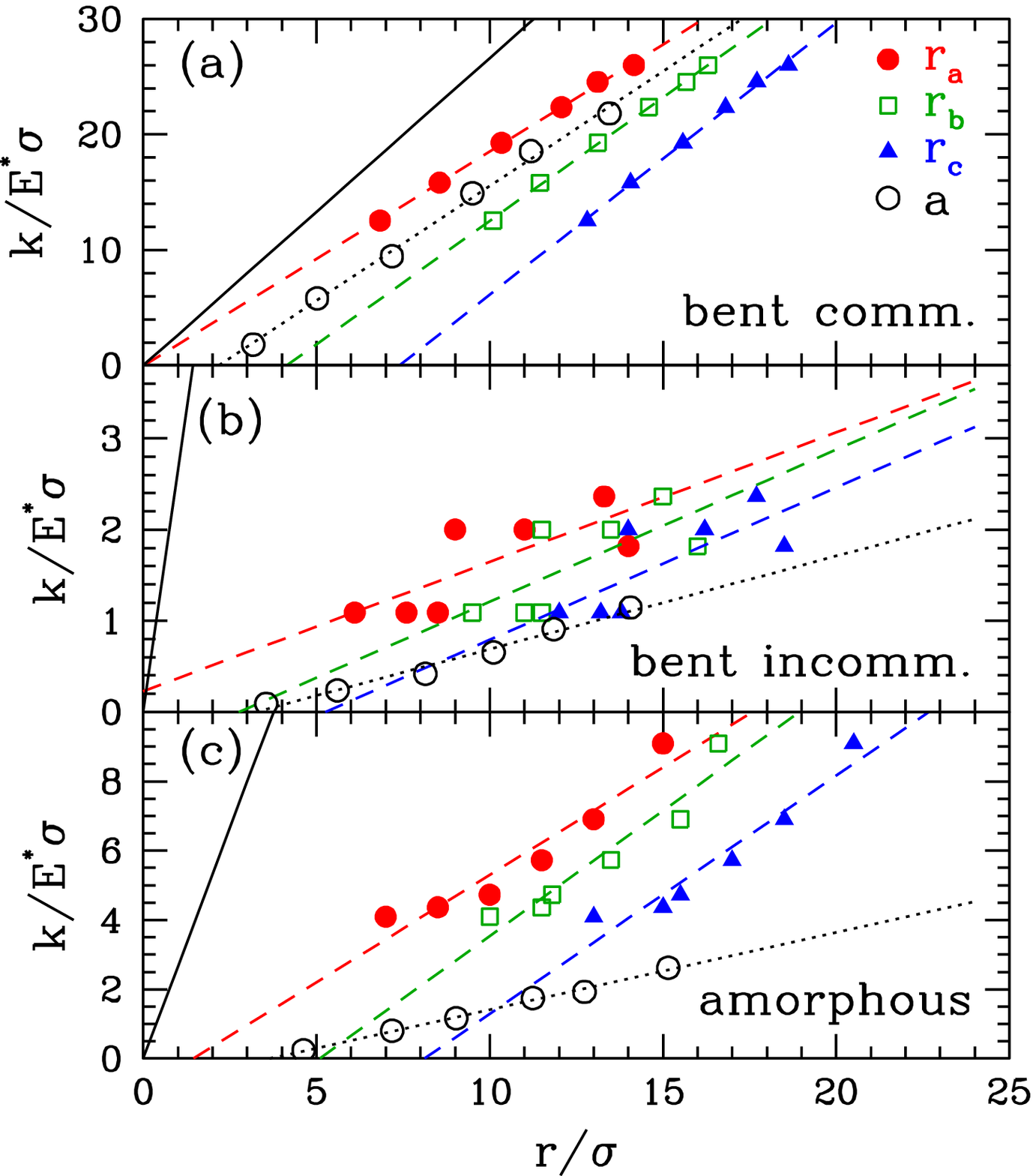}
\caption{
(Color online)
Lateral stiffness $k$ plotted against radius for the indicated tip geometries with same parameters as in Fig. \ref{fig:r05rofN}.
Values of $r_a$ (filled circles), $r_b$ (open squares), and $r_c$ (filled triangles) are shown for adhesive tips.
Open circles show $F$ vs. $a$ for non-adhesive tips.
Broken lines show unconstrained linear fits to each data set,
and the solid line indicates the slope predicted by continuum theory.
Numerical uncertainties are comparable to the symbol size, except for the adhesive data in (b) where they may be up to 50\%.
}
\label{fig:r05kofr}
\end{center}
\end{figure}

Adding adhesion introduces a substantial width to the edge of the contact, ranging from the point where interactions first become attractive $r_a$ to the outer limits of attractive interactions $r_c$ (Fig. \ref{fig:cutoff}).
As the range of interactions increases, it becomes increasingly difficult to fit both these characteristic radii and the pulloff force with the simple M-D theory (Fig. \ref{fig:cutoff22}).
For short-range interactions, good fits are obtained with the measured $w$ for bent tips.
Data for amorphous tips can only be fit by increasing $w$, due to the reduction in effective roughness mentioned above (Fig. \ref{fig:r05rofN}).
For stepped tips the contact radius increases in discrete jumps as successive terraces contact the surface.

The normal interfacial compliance leads to significant ambiguity in the definition of the normal displacement as a function of load (Fig. \ref{fig:r05dofN}).
Continuum theory normally includes only the substrate compliance, while experimental measures of the total tip displacement $\delta_{\rm tip}$ include the interfacial compliance.
The substrate compliance was isolated by following the displacement of substrate atoms, $\delta_{\rm sur}$ and found to agree well with theory for bent tips in the repulsive regime.
Johnson and Greenwood's extension of M-D theory \cite{johnson97} includes the interfacial compliance in the attractive tail of the potential.
It provides a good description of $\delta_{\rm tip}$ in the regime where $r_a=0$.
Here $\delta_{\rm tip} - \delta_{\rm sur}$ increases to the interaction range $h_0$.
Results for amorphous tips show the greater adhesion noted above.
Stepped tips follow continuum theory at large loads but are qualitatively different at negative loads.

The most profound effects of tip geometry are seen in the lateral stiffness $k$ and friction $F$, which vary by one and two orders of magnitude respectively.
Continuum theories for $k$ do not include the lateral interfacial compliance $k_{\rm i}$.
This adds in series with the substrate compliance $k_{\rm sub}$ (Eq. (\ref{eq:stiff})).
Except for commensurate tips, $k_{\rm i} << k_{\rm sub}$ and the interface dominates the total stiffness \cite{luan05}.
Experiments have also seen a substantial reduction in the expected lateral stiffness from this effect \cite{socoliuc04,carpick97thesis}.

The friction on non-adhesive commensurate tips (bent or stepped) increases linearly with load, as frequently observed for macroscopic objects.
In all other cases, $F$ is a nonlinear function of load.
Our ability to directly measure contact radii allowed us to show that $F$ scales linearly with contact area for incommensurate, amorphous and adhesive bent commensurate tips.
These tips also show a linear scaling of $k$ with radius.
While these scalings held for any choice of radius, the linear fits are offset from the origin.
It appears that the effective area contributing to friction and stiffness is often a little smaller than the area of repulsive interactions corresponding to $r_a$.
As noted above, the offset from $r_a$ appears to correspond to the finite range over which repulsive forces rise at the interface.

Experimental data \cite{carpick97,carpick97b,pietrement01,lantz97,carpick04,carpick04b,schwarz97,schwarz97b} for friction and stiffness have been fit to continuum theory with the assumptions that $F \propto r^2$ and $k \propto r$, but without the offsets seen in Figs. \ref{fig:r05Fofr} and \ref{fig:r05kofr}.
We showed that our data for bent and amorphous tips could be fit in this way (Fig. \ref{fig:r05FofN}), but that the fit parameters did not correspond to directly measured values.
This suggests that care should be taken in interpreting data in this manner.

\begin{figure}
\begin{center}
\includegraphics[width=12cm]{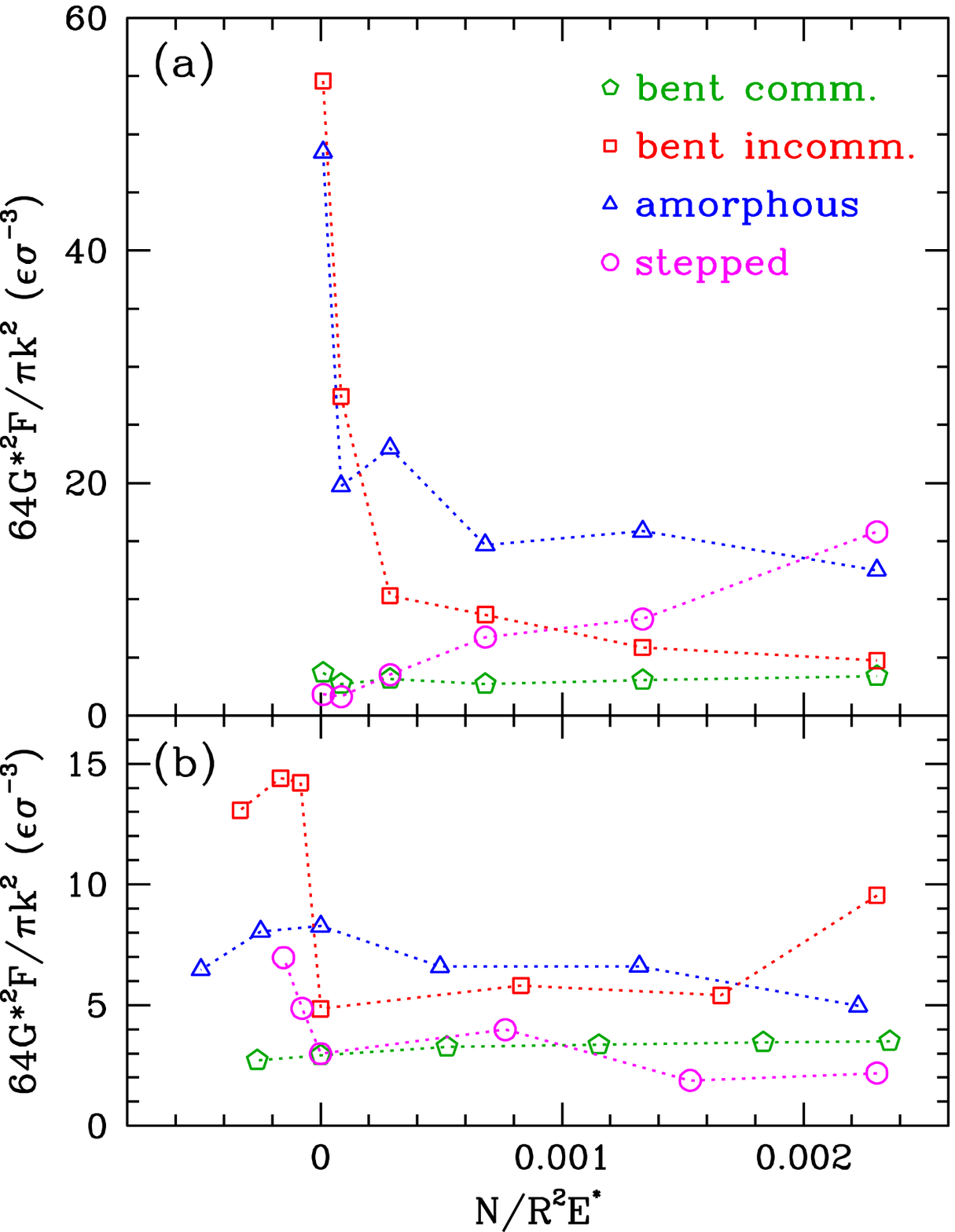}
\caption{
(Color online)
Ratio of friction to stiffness squared as a function of load for the indicated tip geometries (a) without and (b) with adhesion ($w=0.46\ \epsilon/\sigma^2$).
Lines are guides to the eye.
Numerical uncertainties are comparable to the symbol size, except for the bent incommensurate data in (b), where they may be as large as 50\%.
}
\label{fig:r05ratofN}
\end{center}
\end{figure}

We also examined the ratio $F/k^2$.
In continuum theory, this is related to the friction per unit area $\tau_f^{\rm eff}$ through Eq.  (\ref{eq:taueff}).
Our results for $F/k^2$ (Fig. \ref{fig:r05ratofN}) show the range of behaviors observed in experiments, with a relatively constant value for adhesive cases, a rapid increase at low loads in some nonadhesive cases, and a linear rise for non-adhesive stepped tips.
The directly measured values of $\tau_f$ (Table II) are smaller than $\tau_f^{\rm eff}$
by up to two orders of
magnitude, and have qualitatively different trends with tip geometry.
The difference is related to a reduction in the stiffness $k$ 
due to interfacial compliance.
This reduces the inferred value of bulk shear modulus $G'$ and increases the
calculated contact area at any given area.
We expect that experimental results for $F/k^2$ will produce similar overestimates of the true interfacial shear force.

It remains unclear why $F$ and $k$ should follow the observed dependence on $r_a$.
Analytic arguments for clean, flat surfaces indicate that $F$ is very sensitive
to structure, with the forces on commensurate, incommensurate and disordered
surfaces scaling as different powers of area \cite{muser01prl,muser03acp}.
Only when glassy layers are introduced between the surfaces, does the
friction scale in a universal manner \cite{he99,muser01prl,he01tl,he01b}.
Wenning and M\"user \cite{wenning01} have argued that the friction on clean, amorphous tips
rises linearly with area because of a cancellation of two factors,
but have not considered $k$.
Naively, one might expect that the length over which the force rises
is a constant fraction of the lattice spacing and that $k$ is proportional
to $F$.
However, the friction traces change with load and do not always drop
to zero between successive peaks.
We hope that our results will motivate further analytic studies of
this problem, and simulations with glassy films and more realistic potentials.

While we have only considered single asperity contacts in this paper, it is likely that the results are relevant more broadly.
Many experimental surfaces have random roughness on all scales that can be described by self-affine fractal scaling.
Continuum models of contact between such surfaces show that the radius of most of contacts is comparable to the lower length scale cutoff in fractal
scaling \cite{hyun04,pei05}.
This is typically less than a micrometer, suggesting that typical contacts have nanometer scale dimensions where the effects considered here will be relevant.

\acknowledgments

We thank G. G. Adams, R. W. Carpick, K. L. Johnson, M. H. M\"user and I. Sridhar for useful discussions.
This material is based upon work supported by the National Science Foundation under Grants No. DMR-0454947, CMS-0103408, PHY99-07949 and CTS-0320907.

\newpage

\begin{table}
\label{table:work}
\caption
{Relations between interaction strength $\epsilon_i$, work of adhesion $w$, $N_c$ and dimensionless pulloff force $N_c/(\pi w R)$ for tips with different atomic-scale geometries.
The last column gives the dimensionless pulloff force from M-D theory,
$N_c^{M-D}$, with the measured $w$ and $h_0=0.5\ \sigma$.
Values are accurate to the last significant digit.
}
\begin{tabular}{|l|c|c|c|c|c|}
\hline
\hline
Tip geometry & ${\epsilon_i} \over {\epsilon}$ & $w ({\epsilon \over \sigma^2}) $ &$|N_c| ({\epsilon \over \sigma})$& ${{|N_c|}\over{\pi wR}}$  & $ { {|N_c^{M-D}|}\over{\pi w R}}$ \\
\hline
Commensurate & 0.213       & 0.46  &256      & 1.77 &  1.74   \\ \hline
             & 0.5       & 1.05   &569     & 1.72 &  1.64  \\ \hline
Incommensurate & 0.535       & 0.46  &258      & 1.79 &  1.74   \\ \hline
             & 0.5       & 0.45   &238     & 1.68 &  1.74   \\ \hline
Amorphous & 1.0       & 0.46  &326   & 2.26 &  1.74    \\ \hline
          & 0.5       & 0.23  &136      & 1.88 &  1.83   \\ \hline
Stepped & 0.213       & 0.46  &104     & 0.72 &  1.74  \\ \hline
        & 0.5       & 1.05    &168    & 0.51 &  1.64   \\ \hline
\end{tabular}
\end{table}

\begin{table}
\label{table:stress}
\caption
{Frictional stresses $\tau_f$ and apparent moduli $G'$ evaluated from the derivatives of fits in Figs. \ref{fig:r05Fofr} and \ref{fig:r05kofr}, respectively, as a function of tip geometry and work of adhesion $w$.
Values of $G'$ are less than the effective shear modulus of the substrate,
$G^* = 18.3\ \epsilon/\sigma^3$,
and frictional stresses are much smaller than expected from continuum expressions (Fig. \ref{fig:r05ratofN}). 
Statistical errorbars are indicated in parentheses.
}
\begin{tabular}{|l|c|c|c|}
\hline
\hline
Tip geometry & $w (\epsilon \sigma^{-2}) $ &$\tau_f (\epsilon \sigma^{-3})$& $G'(\epsilon \sigma^{-3})$    \\ \hline
Commensurate & 0       & 1.35(7)  & 13.7(4) \\ \hline
             & 0.46       & 1.82(4) & 12.8(3) \\ \hline
Incommensurate & 0       & 0.0044(7) & 0.70(5) \\ \hline
             & 0.46       & 0.0151(8)   & 1.0(3) \\ \hline
Amorphous & 0       & 0.056(4)  &  1.53(5)  \\ \hline
          & 0.46       & 0.24(2)  &  4.3(6)     \\ \hline
\end{tabular}
\end{table}

\begin{thebibliography}{62}
\bibitem[{foo({\natexlab{a}})}]{foot1}
\bibinfo{note}{Note that the geometry of stepped tips is very sensitive to the
  orientation of the crystal before cutting. A slight tilt away from the (001)
  orientation used here leads to inclined steps and a very different pressure
  distribution.}
\bibitem[{foo({\natexlab{b}})}]{footn}
\bibinfo{note}{The tangential force shows smaller fluctuations about zero
  and the radial average is less than 10\% of that for the normal stress.}
\bibitem[{foo({\natexlab{b}})}]{foot2}
\bibinfo{note}{Note that if the friction on amorphous and incommensurate
  surfaces continued scaling as $N^{2/3}$ at low $N$, it would exceed the
  friction on commensurate tips. However, this does not happen until the radius
  is comparable to an atom and their is no distinction between commensurate and
  incommensurate at such small scales.}
\end{thebibliography}
\end{document}